\documentclass[a4paper,fleqn,usenatbib]{mnras}

\usepackage[T1]{fontenc}
\usepackage{ae,aecompl}

\usepackage{graphicx}	
\usepackage{amsmath}	
\usepackage{amssymb}	

\title[Peculiar compact stellar systems in Fornax]{Peculiar compact stellar systems in the Fornax cluster}

\author[C. Wittmann et al.]{
Carolin Wittmann,$^{1}$\thanks{E-mail: cw@x-astro.net}
Thorsten Lisker,$^{1}$
Anna Pasquali,$^{1}$
Michael Hilker$^{2}$
\newauthor
and Eva K. Grebel$^{1}$
\\
$^{1}$Astronomisches Rechen-Institut, Zentrum f\"ur Astronomie der Universit\"at Heidelberg, M\"onchhofstra{\ss}e 12-14, D-69120 Heidelberg, Germany\\
$^{2}$European Southern Observatory, Karl-Schwarzschild-Stra{\ss}e 2, D-85748 Garching, Germany\\
}

\date{Accepted 2016 April 7. Received 2016 March 13; in original form 2015 November 5}

\pubyear{2016}

\begin{document}
\label{firstpage}
\pagerange{\pageref{firstpage}--\pageref{lastpage}}
\maketitle

\begin{abstract}
We search for hints to the origin and nature of compact stellar systems in the magnitude range of ultracompact dwarf galaxies in deep wide-field imaging data of the Fornax cluster core. We visually investigate a large sample of 355 spectroscopically confirmed cluster members with $V$-band equivalent magnitudes brighter than $-10$~mag for faint extended structures. Our data reveal peculiar compact stellar systems, which appear asymmetric or elongated from their outer light distribution. We characterize the structure of our objects by quantifying their core concentration, as well as their outer asymmetry and ellipticity. For the brighter objects of our sample we also investigate their spatial and phase-space distribution within the cluster. We argue that the distorted outer structure alone that is seen for some of our objects, is not sufficient to decide whether these systems have a star cluster or a galaxy origin. However, we find that objects with low core concentration and high asymmetry (or high ellipticity) are primarily located at larger cluster-centric distances as compared to the entire sample. This supports the hypothesis that at least some of these objects may originate from tidally stripped galaxies.
\end{abstract}




\section{Introduction}
\label{sect:sect1}
The discovery of a new type of compact stellar system in the Fornax cluster with luminosities comparable to dwarf galaxies, but of much smaller size, was reported by \citet{hilker:1999} and \citet{drinkwater:2000}. The detected objects, named `ultracompact dwarf galaxies' (UCDs) by \citet{phillipps:2001}, started to fill the mass--size parameter space in between globular clusters (GCs) and compact elliptical galaxies (cEs) in the stellar mass range $M_*$~$=$~$10^6-10^8$~M$_{\sun}$ and with half-light radii between $r_\mathrm{h}$~$=$~$3$ and $100$~pc. With the growing number of UCDs, compact galaxies and star clusters are no longer separated, but form instead one sequence of compact stellar systems. One of the main questions driving the investigation of UCDs is thus whether they rather constitute the high-mass, large-size end of the star cluster distribution, or the low-mass and small-size end of the galaxy population.

Previous studies showed that most UCDs have properties similar to GCs with regard to their ages, metallicities and $\alpha$-element abundances \citep[e.g.][]{frank:2011, francis:2012}. Furthermore, on the basis of number counts, \citet{mieske:2012} concluded that UCDs in the Fornax cluster would statistically be fully consistent with being the brightest members of the central cluster galaxy's GC population. There is an ongoing discussion on the formation scenarios of GCs themselves \citep[e.g.][]{kruijssen:2014, kruijssen:2015}. One model suggests that the formation of GCs may be triggered during major mergers of gas-rich galaxies \citep{ashman:1992}. Support for this scenario can be found in observations of the interacting Antennae galaxies \citep{whitmore:1995}, which show large star cluster complexes residing in giant \ion{H}{ii} regions, consisting of up to hundreds of individual young massive star clusters. Simulations of the evolution of such young massive star clusters in star cluster complexes showed that the merging of clusters would form a compact object with parameters in the range of typical UCDs \citep{fellhauer:2002, bruens:2011}. However, hierarchical merging in star cluster complexes is not the only possible formation scenario for UCD-like objects during a galaxy merger. \citet{renaud:2015} concluded from a hydrodynamical simulation of an Antennae-like galaxy merger that UCD-like objects can also originate from merging of gas clumps during their formation. Based on observations of UCDs in the Perseus cluster, \citet{penny:2012} suggested a star cluster origin for two UCDs with very blue colours, residing in star-forming filaments of the central cluster galaxy.

Since their discovery, it has also been discussed that UCDs could be related to the population of galaxies. For cEs, which are adjacent to UCDs in magnitude and size, it was already suggested by \citet{faber:1973} that these could be the remnant bulges of stripped spiral or early-type galaxies. Observational support for this scenario was found for example by \citet{huxor:2011}. For the lower-mass UCDs it was proposed that they could be the remnant nuclei of nucleated dwarf ellipticals that were tidally stripped while orbiting in the gravitational field of a galaxy cluster \citep{bekki:2001, drinkwater:2003}. Various simulations of this tidal stripping scenario demonstrated that the stripped galaxy remnants would closely resemble observed UCDs from their structural parameters \citep{bekki:2003, goerdt:2008, pfeffer:2013}. Observational signatures for a stripped galaxy origin were found by several authors. For example, \citet{norris:2015} detected an extended star formation history for UCD NGC~4546-UCD1 ($M_*$~$=$~$3.3$~$\times$~$10^7$~M$_{\sun}$, $r_\mathrm{h}$~$=$~$25.5$~pc). Also, \citet{strader:2013} found strong indications for a galaxy origin of UCD M60-UCD1 ($M_{\mathrm{dyn}}$~$=$~$2.0$~$\times$~$10^8$~M$_{\sun}$, $r_\mathrm{h}$~$=$~$24.2$~pc), which harbours a central X-ray source and shows strong indications for the presence of a massive black hole that makes up 15~per~cent of the UCD's mass \citep{seth:2014}. The presence of a massive black hole could also be the cause of the elevated dynamical-to-stellar mass ratio that was inferred for UCD M87-S999 ($M_{\mathrm{dyn}}/M_*$~$=$~$8.2$, $M_*$~$=$~$3.9$~$\times$~$10^6$~M$_{\sun}$, $r_\mathrm{h}$~$=$~$20.9$~pc) \citep{janz:2015}. Another example is UCD NGC~1275-UCD13 ($M_*$~$=$~$4.4$~$\times$~$10^7$~M$_{\sun}$, $r_\mathrm{h}$~$=$~$85$~pc), for which \citet{penny:2014} concluded that the UCD's colour, size, metallicity, internal velocity dispersion, dynamical mass, and close proximity to the central cluster galaxy likely point to a stripped galaxy origin. \citet{janz:2016} observed that high-mass UCDs with $M_*$~$\gtrsim$~$10^7$~M$_{\sun}$ seem to be generally metal-rich in comparison to early-type galaxies in the same stellar mass range. This may indicate that the objects once were more massive galaxies that were stripped off their stellar material while retaining their central metallicity.

In general it is believed that both formation channels coexist for the population of UCDs \citep[e.g.][]{hilker:2009, norris:2014}. For a few UCDs, some of which are mentioned above, either a star cluster or a galaxy origin was found to be more likely. In most cases, however, it was not possible so far to robustly distinguish between the proposed formation scenarios from the observed properties. This is also due to limited resolution and image depth, restricting a detailed analysis of the internal properties mainly to the brighter objects. Furthermore, due to the existence of a luminosity-size relation for the brighter UCDs \citep[e.g.][]{evstigneeva:2008, caso:2013}, with the fainter objects having smaller half-light radii, most of them appear unresolved or only partly resolved in seeing-limited observations.

Already in the early phase of UCD research, \citet{drinkwater:2003} found that the brightest UCD in Fornax (UCD~3) is significantly extended compared to the few other UCDs known at that time. Structural analysis in high resolution {\it HST} imaging of Virgo and Fornax cluster UCDs showed that some of the brightest UCDs are characterized by a two-component surface-brightness profile with a compact core and an extended low-surface-brightness envelope \citep{depropris:2005, evstigneeva:2007, evstigneeva:2008}. In ground-based imaging, a large number of UCDs with faint envelopes were detected in the Virgo cluster \citep{liu:2015}. In the Fornax cluster, \citet{richtler:2005} and \citet{voggel:2016} reported that a few extended objects appear peculiar since they show asymmetric structures.

In deep imaging data of the Fornax cluster core, we search the known population of compact stellar systems for objects that appear significantly extended and/or exhibit peculiar structures. We visually investigate a large sample of 355 spectroscopically confirmed compact systems with $V-$band equivalent magnitudes between $-14$ and $-10$~mag for the presence of faint structures extending significantly beyond the point spread function (PSF). We quantify structure by introducing parameters that measure the core concentration, as well as the outer asymmetry and ellipticity. Then we relate these parameters to the spatial and phase-space distribution within the Fornax cluster. We limit our analysis to objects brighter than $-10$~mag, which corresponds to the magnitude range of UCDs or bright GCs \citep{misgeld:2011, norris:2014}. Since there is no unambiguous distinction of objects that have previously been classified as bright GCs and objects that were named UCDs, we will speak of `compact stellar systems' throughout this paper.

We describe the observations and the sample definition in Section~\ref{sect:sect2}. Details of the analysis are given in Section~\ref{sect:sect3}. We present our results in Section~\ref{sect:sect4} and discuss them in Section~\ref{sect:sect5}. The conclusions follow in Section~\ref{sect:sect6}.

\section{Data}
\label{sect:sect2}

\subsection{Observations}
\label{sect:sect2.1} 
\begin{figure}
	\includegraphics[width=\columnwidth]{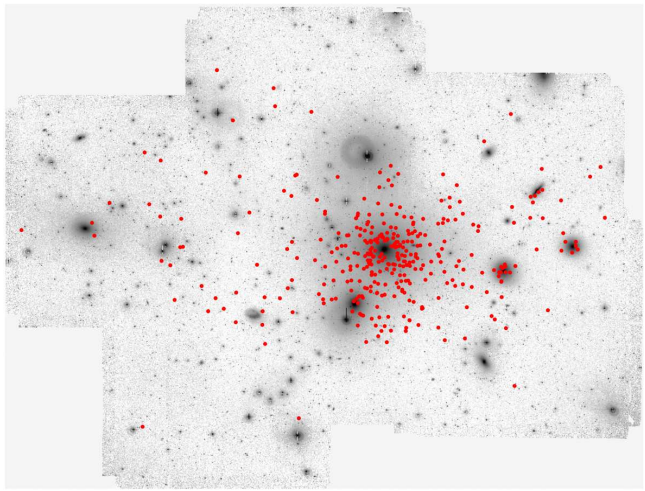}
    \caption{Deep wide-field imaging of the Fornax cluster core. Image dimensions: $100$~$\times$~$76$~arcmin$^2$, corresponding to $576$~$\times$~$438$~kpc$^2$ when assuming a distance of 20.0~Mpc for Fornax. North is up and east is to the left. Red dots indicate our sample of spectroscopically confirmed compact stellar systems with $m_{\mathrm{Ve}}$~$<$~$21.5$~mag. The central galaxy is NGC~1399.}
    \label{fig:fig1}
\end{figure}

The data are fully characterized by \citet{lisker:2016}. They were acquired in 2008 and 2010 with the WFI at the ESO/MPG 2.2m telescope (programmes 082.A-9016 and 084.A-9014, PI A.\ Pasquali, Guaranteed Time of the Max Planck Institute for Astronomy). We used a transparent filter that nearly equals the no-filter throughput ($>10$~per~cent in the range 350--900~nm) and thus provides a high signal-to-noise ratio. Fig.~\ref{fig:fig1} shows our deep image of the Fornax cluster core, covering a region of 100~arcmin in east--west and 76~arcmin in north--south direction. This corresponds to 576 and 438~kpc, respectively, at a distance of 20.0 Mpc for Fornax (using 5.76~kpc arcmin$^{-1}$; \citealt{blakeslee:2009}), which we adopt throughout this work.

\citet{lisker:2016} determined an approximate `$V-$equivalent' magnitude calibration, based on spectroscopically confirmed foreground stars from \citet{FCOS, FCOS2}. In the following we denote $V-$equivalent magnitudes and surface brightnesses with the subscript `${Ve}$'. Our data reach a $V-$equivalent depth of 26.58 and 26.76~mag~arcsec$^{-2}$ for the median and 75th percentile, respectively, at S/N $= 1$ per pixel (1 pixel $\cor 0.238$~arcsec, or $22.848$~pc). The seeing PSF full width at half-maximum (FWHM) varies over the image (see also Fig.~\ref{fig:fig2}), but is typically about 1~arcsec. Details of the observations, data reduction, and data characterization are provided in \citet{lisker:2016}.

\subsection{Catalogue compilation}
\label{sect:sect2.2}

Before defining our working sample, we first compiled a list of published compact stellar systems with spectroscopically confirmed Fornax cluster membership. We used the SIMBAD data base \citep{SIMBAD} to identify relevant source catalogues, which yielded the following references:
\citet{SRH2010}, \citet{DRG2004}, \citet{BAL2007}, \citet{FDE2007}, \citet{GDE2009}, \citet[][part of the Fornax Compact Object Survey]{FCOS2}, \citet[][part of the Fornax Compact Object Survey]{FCOS}, \citet{FDK2008}, \citet[][part of the Fornax Cluster Spectroscopic Survey]{DJG2000}, \citet{KGM1999}, and \citet{KBS1998}. 
Tabulated cluster members from these references, in the given order, were appended to our object list if they were not within 1 arcsec of objects already in the list. Duplications within the same catalogue, which are present in \citet{SRH2010} and \citet{DRG2004}, were excluded. Based on this sequence, 695 objects were taken from \citet{SRH2010}, 113 from \citet{DRG2004}, 151 from \citet{BAL2007}, and 123 from the remaining references (with no contributed objects from \citealt{DJG2000} and \citealt{KBS1998}), resulting in 1082 confirmed cluster members, of which 1058 were listed as compact systems and 24 as dwarf galaxies.

After excluding eight objects that lie outside of our mosaic, we visually inspected the remaining objects on the mosaic at their literature positions, using {\it SAOImage DS9}. We identified another 75 unambiguous duplications, as well as 20 published object coordinates for which we do not see an optical counterpart in our image within 3 arcsec\footnote{These are the objects 78:103, GS04-M03:9, GS04-M03:30, and GS04-M03:127 from \citet{SRH2010}, objects 75:56, 76:112, 78:13, and 78:110 from \citet{DRG2004}, and the following objects from \citet{KGM1999}: ntt 201, ntt 203, ntt 407, ntt 410, ntt 414,
ntt 109, ntt 119, ntt 122, ntt 123, ntt 124, ntt 126, and ntt 127. We used the NED Coordinate and Extinction Calculator (http://ned.ipac.caltech.edu/forms/calculator.html) as a check to make sure that our coordinate conversion from the B1950.0 values of
\citeauthor{KGM1999} to J2000.0 was done correctly.}. For 50 further published object coordinates, the match to the visible sources is ambiguous --- mostly due to the presence of multiple sources, but in a few cases also due to the fact that the only nearby visible source lies almost 3 arcsec away from the literature position. All of these published objects were excluded from our list, as well as one further object that potentially has wrong published coordinates\footnote{Object 37 in table 2 of \citet{GDE2009} is marked there as being the same source as object gc212.2 from \citet{BAL2007}, but that object actually has different coordinates, offset by about 0.5 arcmin. 
We therefore excluded object 37 from our list.}. This results in a final catalogue of 904 compact stellar systems that are spectroscopically confirmed cluster members, are covered by our mosaic, and are visually unambiguously identified with an optical source. The catalogue is available online. An excerpt is shown in Table~\ref{tab:tab1}.

Despite the visually unique match, the literature position may of course still be slightly offset from the image position\footnote{The image position is based on the astrometric calibration of our mosaic (using the 2MASS catalogue), and obtained by fitting the PSF of the objects using the task {\it allstar} of the \texttt{IRAF} package \texttt{DAOPHOT}.}, and other neighbouring sources may be present. In order to characterize this situation, we determined for each catalogue position whether a single source or multiple image sources are present within a 1~arcsec and 3~arcsec radius. For 792 catalogue positions (88\%) there is only a single image source within 1~arcsec and no further image source within 3~arcsec. For 95 catalogue positions (11\%) there is a single image source within 1~arcsec, and at least one further source within 3~arcsec. For 17 catalogue positions (2\%) there is no image source within 1~arcsec, but either a single image source within 3~arcsec (15 objects) or multiple sources (2 objects\footnote{The two objects are not part of our working sample, since they have $m_{\mathrm{Ve}}$~$\geq$~$21.5$~mag.}). We remind that a \emph{visually unambiguous} match is present for \emph{all} 904 catalogue objects, including the 17 just mentioned.

\begin{table*}
 \caption{Catalogue of compact stellar systems compiled from the literature. The first five objects are given to illustrate the format of the table. The complete catalogue is provided in the electronic version of the paper. For each object we list our ID, the position based on our astrometry, as well as the SIMBAD identifier. The object IDs are sorted by increasing right ascension. The given velocity corresponds to the velocity with the smallest error from all compiled velocities. The respective literature source is listed in the last column, where 1 = \citet{SRH2010}, 2 = \citet{DRG2004}, 3 = \citet{BAL2007}, 4 = \citet{FDE2007}, 5 = \citet{GDE2009}, 6 = \citet{FCOS2}, 7 = \citet{FCOS}, 8 = \citet{FDK2008}, 9 = \citet{KGM1999}, 10 = \citet{DJG2000}, 11 = \citet{KBS1998}.}
  \label{tab:tab1}
 \begin{tabular}{ccccccc}
  \hline
  ID & R.A. (J2000) & Dec. (J2000) & SIMBAD ID & v (km s$^{-1}$) & Lit.\\
   \hline
  1 & 03 35 38.87 & -35 21 53.3 & [BAL2007] gc144.6 & 1388 $\pm$ 32 & 3\\
  2 & 03 35 42.52 & -35 13 51.8 & [BAL2007] gc290.6 & 1901 $\pm$ 16 & 3\\
  3 & 03 35 50.49 & -35 15 24.2 & [BAL2007] gc302.6 & 1166 $\pm$ \phantom{0}6 & 3\\
  4 & 03 35 59.56 & -35 26 56.7 & [BAL2007] gc21.70 & 1272 $\pm$ 12 & 3\\
  5 & 03 36 01.09 & -35 25 43.0 & [BAL2007] gc69.70 & 1389 $\pm$ \phantom{0}8 & 3\\
  \hline
 \end{tabular}
\end{table*}

\subsection{Velocity compilation}
\label{sect:sect2.3}
In the literature compilation described in Section~\ref{sect:sect2.2} that serves as the basis for our sample, each object only appears once, even if it is listed in several of the references. In order to compile \emph{all} published heliocentric velocity measurements of a given object, we employed the following steps. First, we went through the list of compiled \emph{literature positions} of our sample and extracted the nearest matching object within 1 arcsec from each of the literature references given in Section~\ref{sect:sect2.2}. Secondly, we repeated this for the list of \emph{image positions} of our sample. Thirdly, we visually inspected the location of all compiled positions, and only kept the unambiguous ones.

This resulted in 440 objects of our sample having at least two velocity measurements, 80 having at least three, and 37 with four or more. 
For each object we adopted the velocity with the smallest error from all compiled velocities, which we include in our catalogue (see Table~\ref{tab:tab1}).

\subsection{Working sample}
\label{sect:sect2.4}
The compiled catalogue contains objects in both the UCD and GC magnitude range. In the study presented here we define our working sample to include all objects brighter than $m_{\mathrm{Ve}} = 21.5$~mag, corresponding to $M_{\mathrm{Ve}} < -10.0$~mag at the distance of Fornax, i.e. the range of UCDs and bright GCs in the literature \citep{misgeld:2011, norris:2014}. Due to the depth and seeing-limited resolution of our data, fainter objects (with their smaller intrinsic sizes, see e.g. \citealt{caso:2013}) could not be analysed robustly. In total our working sample contains 355 compact systems, which are indicated in Fig.~\ref{fig:fig1}. Their $V$-equivalent magnitudes (`mag best') and the corresponding uncertainties (based on the weight image) were obtained with \texttt{SExtractor}\footnote{For five objects from our working sample we adopted the magnitude and corresponding uncertainty obtained from PSF fitting, due to imperfect deblending in \texttt{SExtractor}.} \citep{SExtractor} and are provided in Table~\ref{tab:tab3}.

A rough estimate of the completeness of known compact stellar systems with $M_{V} < -10.3$~mag in Fornax is given by \citet{mieske:2012}. Between distances of 50--100~kpc from the cluster centre (corresponding to $0.15\degr$ and $0.29\degr$, respectively) the authors considered their sample to be complete to 60--70 per cent. At larger cluster-centric distances the completeness is expected to drop below 50 per cent. As discussed in Section~\ref{sect:sect5.3} this does not affect the conclusions of our study.

\section{Analysis}
\label{sect:sect3}
The analysis of our working sample is based on both a visual investigation as well as on a parametrization of selected sample properties. We introduce parameters as a measure for the central core concentration as well as for the shape of the outer light distribution. The analysis is carried out relative to the PSF, which is determined from point sources in our data. Prior to the analysis we have removed large galaxies and stellar haloes by fitting and subtracting their light profile with the \texttt{IRAF} task \textit{ellipse} \citep[see][]{lisker:2016}.

\subsection{PSF analysis and background correction}
\label{sect:sect3.1}

\begin{figure}
	\includegraphics[width=\columnwidth]{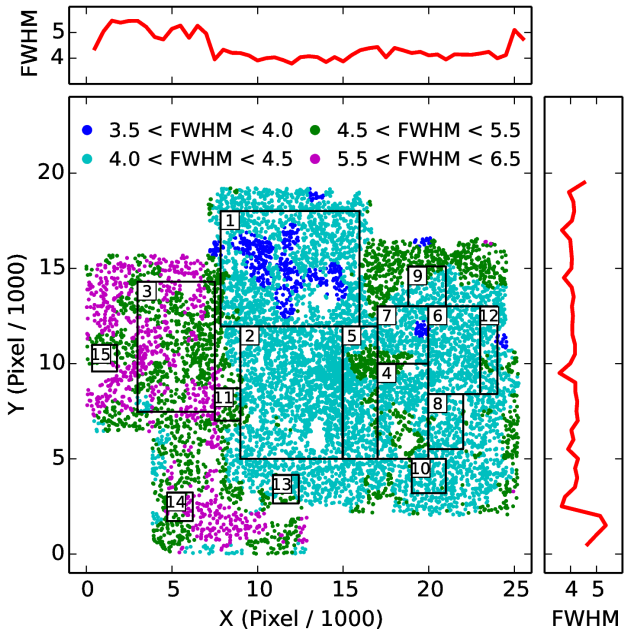}
    \caption{Central panel: variation of the PSF FWHM across the mosaic. The positions of objects brighter than $m_{\mathrm{Ve}} = 22$~mag and with an FWHM below 7.5 pixels are shown, which were detected with \texttt{SExtractor}. The symbols are colour-coded according to the local minimum FWHM, determined within a circle of 500 pixels radius. We measured the FWHM of all \texttt{SExtractor}-detected objects with the \texttt{IRAF} task {\it psfmeasure}, using a Gaussian profile for the PSF. The numbered boxes indicate regions for which we determined an individual PSF. North is up and east is to the left. Side panels: variation of the minimum FWHM along $x$-$\mathrm{/}y$-direction. The average of the 10 smallest FWHM values that occur in bins of 500 pixels is plotted.}
    \label{fig:fig2}
\end{figure}

Due to changing seeing conditions over the course of our observing runs, aside from possible intrinsic variations due to the instrument, the PSF varies across the image. We therefore divided the mosaic into regions of similar FWHM, as indicated in Fig.~\ref{fig:fig2}, and according to the distribution of our sample, i.e., only if an area contains published compact stellar systems, we define it as a PSF region. In each region we determined the PSF from suitable point sources (hereafter `PSF stars') in the magnitude range $17.6$~$<$~$m_{\mathrm{Ve}}$~$<$~$22.3$~mag, using routines from the \texttt{IRAF} package \texttt{DAOPHOT}. The PSF was generated by an iterative approach. In a wide circle with a radius of 30~pixels (7.14~arcsec) around each PSF star, all neighbouring sources were subtracted using a first-estimate PSF. A new PSF model was then calculated based on the neighbour-subtracted PSF stars. We performed three iterations, which significantly reduced the contamination of flux from close neighbours to the PSF. The final PSF was subsequently fitted to our sample and subtracted, where we chose the fitting radius to be slightly larger than the typical FWHM of the respective region. Details on the PSF analysis and properties of the resulting PSFs are summarized in Table~\ref{tab:tab2}.

For the following analysis we corrected large-scale background variations using a \texttt{SExtractor} background map determined from the full image. Remaining local background offsets were corrected for each object individually, by subtracting the median intensity within an annulus 5~pixels wide and with an inner radius of 20~pixels (1 pixel $\cor 0.238$~arcsec).

\begin{table*}
 \caption{PSF analysis for regions of similar FWHM on the mosaic (see Fig.~\ref{fig:fig2}). For each region we specify the size of the region in pixels$^2$, the number of compact stellar systems contained in the region, and the number of PSF stars used to build the PSF. The regions are listed with decreasing size. Working sample: compact objects brighter than $m_{\mathrm{Ve}} = 21.5$~mag.  Fainter CSS: compact objects fainter than $m_{\mathrm{Ve}} = 21.5$~mag, which belong to the basis catalogue described in Section~\ref{sect:sect2.2} and are given here for completeness only. For the resulting PSF we give the FWHM in pixels (1 pixel $\cor 0.238$~arcsec) and the ellipticity. Both quantities were obtained with the \texttt{IRAF} task {\it psfmeasure}, using a Gaussian profile. The last column gives the mean core concentration ($\overline{\mathrm{cc}}$) and the corresponding standard deviation of all PSF stars within one region (see Section~\ref{sect:sect3.2}).}
  \label{tab:tab2}
 \begin{tabular}{lccccccc}
  \hline
  Region & Size & Working sample & Fainter CSS & PSF stars & PSF FWHM & PSF ellipticity & $\overline{\mathrm{cc}}$\\
  \hline
  \phantom{0}1 & 8100 $\times$ 6050 & \phantom{0}16 & \phantom{00}4 & 78 & 4.1 & 0.09 & 1.005 $\pm$ 0.025\\
  \phantom{0}2 & 5980 $\times$ 6950 & 132 & 223 & 55 & 4.3 & 0.05 & 1.000 $\pm$ 0.019\\
  \phantom{0}3 & 4500 $\times$ 6840 & \phantom{0}13 & \phantom{0}10 & 42 & 5.6 & 0.10 & 1.006 $\pm$ 0.027\\
  \phantom{0}4 & 2970 $\times$ 4980 & \phantom{0}37 & \phantom{0}40 & 28 & 4.6 & 0.04 & 1.012 $\pm$ 0.031\\
  \phantom{0}5 & 2040 $\times$ 6950 & 107 & 246 & 35 & 4.5 & 0.05 & 1.000 $\pm$ 0.037 \\
  \phantom{0}6 & 3030 $\times$ 4600 & \phantom{0}21 & \phantom{00}2 & 37 & 4.4 & 0.05 & 1.006 $\pm$ 0.029\\
  \phantom{0}7 & 2970 $\times$ 3020 & \phantom{0}15 & \phantom{0}24 & 22 & 4.4 & 0.02 & 1.013 $\pm$ 0.042\\
  \phantom{0}8 & 2030 $\times$ 2900 & \phantom{00}1 & \phantom{00}0 & 28 & 4.5 & 0.07 & 1.008 $\pm$ 0.026\\
  \phantom{0}9 & 2200 $\times$ 2100 & \phantom{00}2 & \phantom{00}0 & \phantom{0}8 & 4.4 & 0.05 & 1.007 $\pm$ 0.018\\
  10 & 2000 $\times$ 1800 & \phantom{00}1 & \phantom{00}0 & \phantom{0}9 & 4.4 & 0.12 & 1.006 $\pm$ 0.022\\
  11 & 1500 $\times$ 1700 & \phantom{00}4 & \phantom{00}0 & \phantom{0}7 & 4.9 & 0.12 & 1.012 $\pm$ 0.040\\
  12 & 1000 $\times$ 2500 & \phantom{00}3 & \phantom{00}0 & 12 & 4.3 & 0.13 & 0.999 $\pm$ 0.020\\
  13 & 1500 $\times$ 1500 & \phantom{00}1 & \phantom{00}0 & \phantom{0}8 & 4.7 & 0.10 & 1.011 $\pm$ 0.024\\
  14 & 1500 $\times$ 1500 & \phantom{00}1 & \phantom{00}0 & 10 & 5.8 & 0.12 & 1.001 $\pm$ 0.031\\
  15 & 1500 $\times$ 1500 & \phantom{00}1 & \phantom{00}0 & \phantom{0}8 & 6.2 & 0.11 & 1.027 $\pm$ 0.049\\
  \hline
 \end{tabular}
\end{table*}

\subsection{Core concentration}
\label{sect:sect3.2}

The core concentration is defined as the mean central flux ratio of the object to the fitted PSF and can be expressed in terms of the ratio of the PSF-subtracted residual image to the object image:
\begin{eqnarray}
\label{eq:eq1}
\text{Core concentration} &=& \frac{1}{n_{\mathrm{pix}}} \sum_{r\leq2\mathrm{pix}} \frac{I_{\mathrm{obj}}}{I_{\mathrm{PSF}}}\\
                          &=& \frac{1}{n_{\mathrm{pix}}} \sum_{r\leq2\mathrm{pix}} \left[ 1- \frac{I_{\mathrm{res}}}{I_{\mathrm{obj}}} \right] ^{-1}\nonumber,
\end{eqnarray}
where $I_{\mathrm{obj}}$ corresponds to the object intensity, $I_{\mathrm{PSF}}$ to the intensity of the fitted PSF, $I_{\mathrm{res}}$ to the residual intensity and $n_{\mathrm{pix}}$ is the number of pixels within the aperture over which is summed. The aperture width is on the order of one FWHM of the typical PSF. The ratio of the residual to the object image was obtained by dividing both images pixel by pixel. We provide the core concentration for our working sample in Table~\ref{tab:tab3}.

\begin{figure*}
	\includegraphics[width=\textwidth]{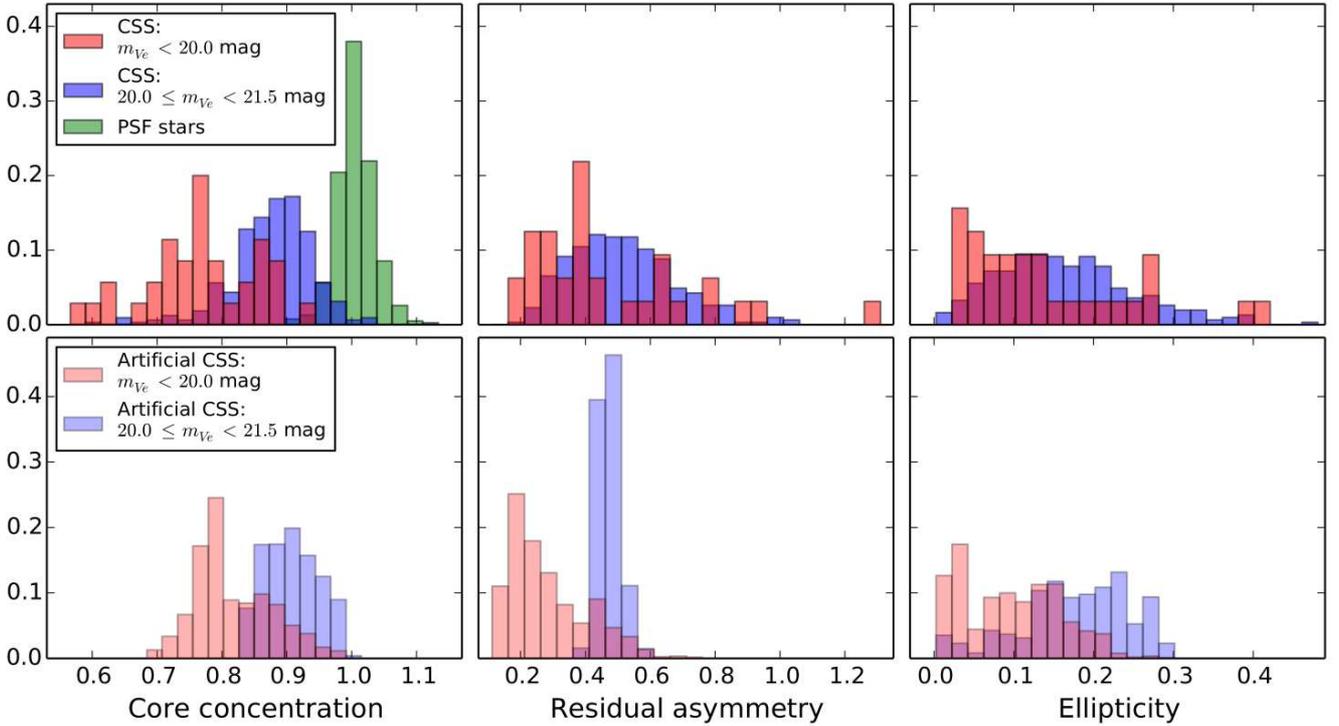}
    \caption{Distribution of parameter values for compact stellar systems and PSF stars. Our working sample corresponds to the red and blue histograms in the upper panels. For comparison we show the distributions of the artificial compact objects in the lower panels, which are represented by the red and blue histograms with lighter shading. Darker shading is used where histograms overlap. In both core concentration histograms we plot the PSF-corrected values (see Section~\ref{sect:sect3.4}). Our sample with $m_{\mathrm{Ve}} < 20.0$~mag contains 35 objects, the sample with $20.0 \leq m_{\mathrm{Ve}} < 21.5$~mag includes 320 objects. For three objects from the brighter and fourteen objects from the fainter sample no useful asymmetry value could be determined due to too little flux remaining in the residual image. The total number of artificial objects corresponds to 1170 for objects with $m_{\mathrm{Ve}} < 20.0$~mag, and to 1215 for objects with $20.0 \leq m_{\mathrm{Ve}} < 21.5$~mag. We note that although both real and artificial compact objects span the same magnitude range, the magnitude distribution of the artificial objects is discrete, since the models were created on the basis of magnitudes and sizes from only 18 real UCDs. Therefore the parameter distributions of the artificial objects should be used for guidance only.}
    \label{fig:fig3}
\end{figure*}

Fig.~\ref{fig:fig3} (upper left panel) shows the distribution of the measured core concentration for our working sample and for the PSF stars used in the analysis. According to equation~\ref{eq:eq1}, a core concentration close to one indicates a PSF-like core, whereas lower values correspond to a less concentrated core. This is reflected in the distribution of the PSF stars, which is strongly peaked around a value of one. The vast majority of compact stellar systems has lower core concentrations than the PSF stars and follows a broader distribution that is skewed towards lower values. We find that the brighter objects from our working sample reach on average lower core concentrations than the fainter objects. This at least partly reflects the luminosity-size relation that exists for the brighter compact stellar systems, such that fainter systems appear PSF-like since they are unresolved in seeing-limited observations.

Due to the high S/N of the central pixels, the formal errors in core concentration are very small when considering uncertainties from photon statistics only\footnote{The relative errors are on the order of 0.04 per cent for objects brighter than $m_{\mathrm{Ve}} = 20.0$~mag, and 0.13 per cent for fainter objects from our working sample.}. An estimate for the (more relevant) uncertainty that stems from PSF fitting is given by the standard deviation of the core concentration of all PSF stars within one region (provided in column~8 of Table~\ref{tab:tab2}). This reflects the local FWHM variations as compared to the PSF model of the respective region.

\begin{table*}
 \caption{Parameter catalogue for our working sample of spectroscopically confirmed Fornax cluster members with $m_{\mathrm{Ve}} < 21.5$~mag. The first five objects are printed below to illustrate the format of the table. We provide the complete table in the electronic version of the paper. ID: our object ID from Table~\ref{tab:tab1}. $m_{\mathrm{Ve}}$: $V$-equivalent magnitude and corresponding uncertainty. Note that the magnitude uncertainties are purely based on S/N and do not include the calibration uncertainties mentioned in Section~\ref{sect:sect2.1}. Flag: 1 = magnitude and uncertainty obtained from \texttt{SExtractor}; 2 = magnitude and uncertainty obtained from PSF-fitting. cc: core concentration (uncorrected values, not used in the analysis). cc$_{\mathrm{corr}}$: PSF-corrected core concentration (see Section~\ref{sect:sect3.4}). An estimate for the uncertainty in core concentration, which is given for each region, can be inferred from column 8 of Table~\ref{tab:tab2}. ra: residual asymmetry and corresponding uncertainty. el: ellipticity and corresponding uncertainty. A value of --99 for ellipticity or residual asymmetry denotes that it was not possible to determine the respective quantity. Sub.: subsample as defined in Table~\ref{tab:tab5};  cc+ra=1, cc+RA=2, CC+RA=3, cc+EL=4; Sub.=0 indicates that the object is not part of any subsample. Alt. sub.: alternative subsample as defined in Table~\ref{tab:tab8}. Region: PSF-region (see Table~\ref{tab:tab2}).}
  \label{tab:tab3}
 \begin{tabular}{cccccccccc}
  \hline
  ID & $m_{\mathrm{Ve}}$ (mag) & Flag & cc & cc$_{\mathrm{corr}}$ & ra & el & Sub. & Alt. sub. & Region\\
   \hline
  1 & 20.394 $\pm$ 0.003 & 1 & 0.946 & 0.946 & 0.504 $\pm$ 0.094 & 0.161 $\pm$ 0.067 & 0 & 0 & 12\\
  2 & 20.576 $\pm$ 0.004 & 1 & 0.929 & 0.929 & 0.429 $\pm$ 0.114 & 0.129 $\pm$ 0.097 & 0 & 0 & 12\\
  3 & 21.379 $\pm$ 0.008 & 1 & 0.908 & 0.908 & 0.403 $\pm$ 0.125 & 0.066 $\pm$ 0.078 & 0 & 0 & 12\\
  4 & 20.737 $\pm$ 0.003 & 1 & 0.942 & 0.942 & 0.520 $\pm$ 0.101 & 0.325 $\pm$ 0.084 & 0 & 0 & \phantom{0}6\\
  5 & 20.645 $\pm$ 0.004 & 1 & 0.905 & 0.905 & 0.759 $\pm$ 0.044 & 0.308 $\pm$ 0.028 & 0 & 0 & \phantom{0}6\\
  \hline
 \end{tabular}
\end{table*}

\subsection{Residual asymmetry and ellipticity}
\label{sect:sect3.3}
For measuring residual asymmetry and ellipticity, we chose to use the PSF-subtracted residual image with all negative pixels set to zero, to avoid effects from the central oversubtraction of the PSF core (see Fig.~\ref{fig:fig4}, left-hand panels), as well as noise effects. For simplicity, we refer to this image as the \emph{residual image} in this subsection.\\
The residual asymmetry is defined in analogy to the Asymmetry parameter in \citet{conselice:2000}:
\begin{equation}
\text{Residual asymmetry} = \frac{\sum_{r\leq12\mathrm{pix}}\left| I_{\mathrm{res,corr}} - I_{\mathrm{res,corr,180}} \right|}{\sum_{r\leq12\mathrm{pix}} I_{\mathrm{res}}}
\label{eq:eq2}
\end{equation}
In order to avoid substantial noise effects due to the small number of pixels (as compared to galaxies), we consider only the flux that exceeds $+1\sigma$ of the noise level\footnote{The noise level of the deeper regions of the mosaic corresponds to $\umu_{\mathrm{Ve}}=26.76$\,mag\,arcsec$^{-2}$ or 0.023 counts, see \citet{lisker:2016}.}: $I_{\mathrm{res,corr}}$ is the intensity on the residual image minus the noise level of 0.023 counts, and is set to zero if this is negative. $I_{\mathrm{res,corr,180}}$ is the corresponding intensity on the image rotated by 180\degr. The sum of the absolute pixel-by-pixel differences of the original and rotated image is then normalized by the flux of the residual image. We use all pixels within a circular aperture with a radius of 12 pixels ($> 2$ PSF FWHM) around the object centre. Any residual light from extended structures is essentially unaffected by the PSF at a radius larger than 2 FWHM; thus a fixed radius ensures comparability of such structures across the mosaic. All close neighbouring sources (and their counterparts rotated by 180\degr) are masked, i.e., their pixels are not taken into account in the asymmetry calculation. 

The ellipticity is defined as $1 - b/a$, where $b/a$ denotes the axis ratio. The latter is computed on the residual image from the second-order moments of the intensity distribution (as done in \texttt{SExtractor}), using the above aperture and neighbour masks.

We estimated the uncertainty in residual asymmetry and ellipticity through error propagation, using the noise level at a given object's position (provided by the weight image) as uncertainty of a pixel's flux. The derived parameter values and corresponding uncertainties are given in Table~\ref{tab:tab3}. We display the distribution of both residual asymmetry and ellipticity in Fig.~\ref{fig:fig3} (upper centre and upper right panels).  The relations between core concentration, residual asymmetry and ellipticity are shown in Appendix~\ref{sect:sectA}.

\subsection{Comparison to artificial compact objects}
\label{sect:sect3.4}
We estimated which parameter values can arise due to the shape of the PSF or the profile type of the objects by comparing the parameters of our sample to the parameter range of artificially created compact objects in the same magnitude and size range. We realized the artificial objects using structural parameters of real Fornax and Virgo cluster UCDs from \citet{evstigneeva:2008}, where we selected 18 one-component UCDs, which have $V-$band magnitudes between $20.90$ and $18.33$~mag and effective radii between $4.0$ and $29.5$~pc. In addition to the observed structural parameters we created further models with different S\'{e}rsic indices ranging from $n = 1$ to $8$. We generated each artificial model with a 10 times smaller pixel scale (i.e., 10 times better sampling) than our actual data, and then convolved it with each of our PSFs. This resulted in a set of roughly 2400 artificial objects for which we determined the core concentration, residual asymmetry, and ellipticity in the same way as for the real objects.

For the artificial objects we see a dependence of the core concentration (cc) on the PSF for some of our PSF models. Basically, the more extended (less concentrated) an object's core is intrinsically, the more the cc value gets lowered by the PSF convolution. This effect is enhanced when the PSF is broad, as compared to a reference PSF that is particularly symmetric and peaked. We chose the PSF from {\it Region~2} (see Table~\ref{tab:tab2} and  Fig.~\ref{fig:fig2}) as reference PSF. For those PSFs showing the strongest deviations, the ratio of the cc value obtained with the reference PSF to the cc value obtained with the actual PSF depends roughly linearly on the latter. Therefore, we can determine correction factors, which need to be applied to the real objects to make their core concentration nearly independent of the PSF, and thus comparable to each other. Corrections were only applied for cc values lower than 0.95 and for five different PSFs ({\it Regions 7, 9, 10, 11} and {\it 13}, see Table~\ref{tab:tab2}), according to:

\begin{equation}
\begin{split}
\mathrm{cc}_{\mathrm{corr}} = \mathrm{cc} \Big[1 + k(\mathrm{PSF_i})~\left(\mathrm{cc}-t\right)\Big],\\t=\begin{cases} 1.0 & \text{{\it Regions 10, 13}}\\0.95&\text{{\it Regions 7, 9, 11,}}\end{cases}
\end{split}
\label{eq:eq3}
\end{equation}
where $k(\mathrm{PSF_i})$ is the PSF-dependent correction factor. In total we corrected cc values of 22 objects from our working sample. The difference between the corrected and uncorrected values never exceeds 10 per cent. We include the corrected core concentration in Table~\ref{tab:tab3}.

We display the distribution of parameter values for the artificial compact objects in the lower panels of Fig.~\ref{fig:fig3}. These show magnitude-dependent differences in their parameter range: brighter objects can reach lower core concentration, whereas fainter objects have in general somewhat higher residual asymmetry and ellipticity. As a consequence, we apply slightly different parameter cuts to our working sample for objects with $m_{\mathrm{Ve}} < 20.0$~mag and for objects with  $20.0$~$\leq$~$m_{\mathrm{Ve}}$~$<$~$21.5$~mag when defining subsamples in the following analysis.

\section{Results}
\label{sect:sect4}
\subsection{Peculiar compact stellar systems}
\label{sect:sect4.1}

We visually investigated compact stellar systems with high residual asymmetry and ellipticity as compared to the artificial compact objects. From our sample with $m_{\mathrm{Ve}}~<~20.0$~mag, we examined those objects with residual asymmetry higher than 0.5 or ellipticity higher than 0.2, corresponding to 34 or 25 per cent of all objects with a measured residual asymmetry parameter in that magnitude range. The overlap with the artificial objects that have residual asymmetries in this parameter range is 6 per cent. The overlap with regard to ellipticity is 5 per cent. From our sample with $20.0 \leq m_{\mathrm{Ve}} < 21.5$~mag, we looked through all objects with a residual asymmetry higher than 0.55, corresponding to 38 per cent, or ellipticity higher than 0.25, corresponding to 16 per cent of compact stellar systems within that magnitude range. The overlap fractions with the artificial objects are 3 and 15 per cent, respectively.

\begin{figure*}
	\includegraphics[width=\textwidth]{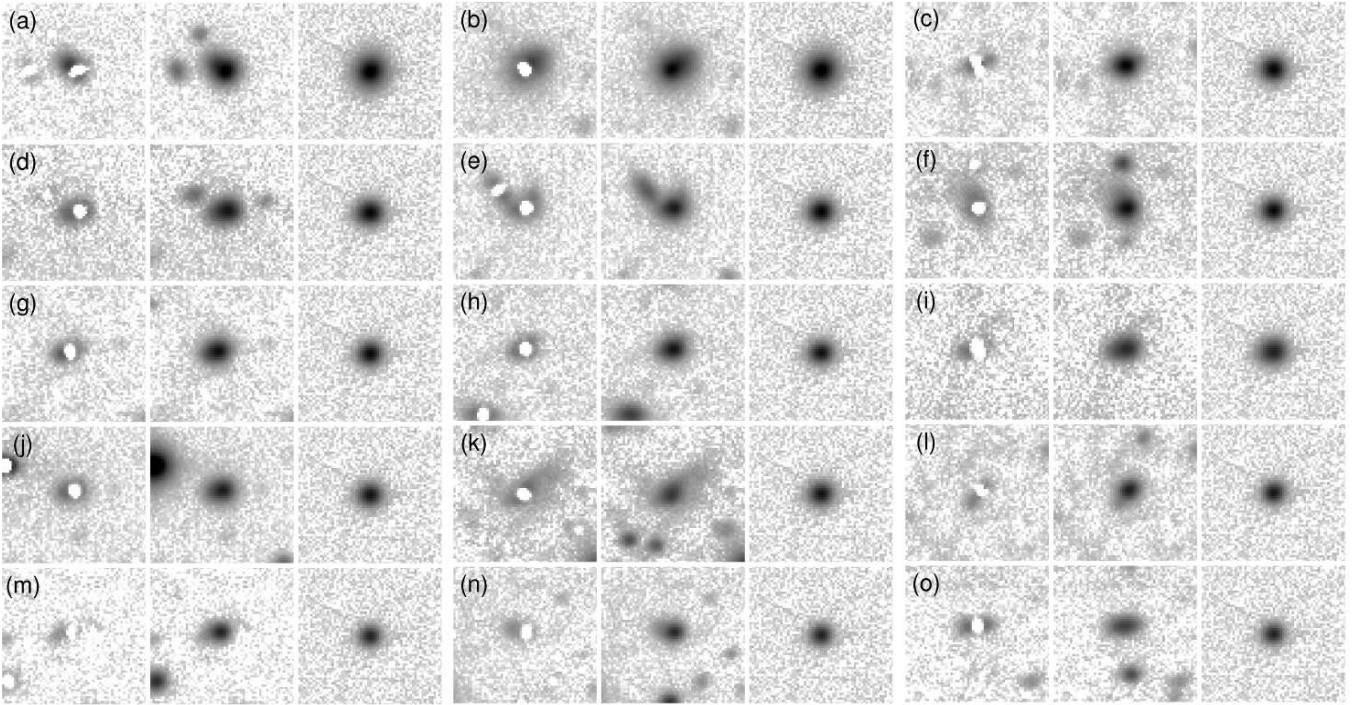}
\caption{Peculiar compact stellar systems from our sample (centre sub-panels). The PSF-subtracted residual images are displayed in the left-hand sub-panels. We compare each peculiar object to an artificial compact object of similar brightness that has been convolved with the PSF of the corresponding region (shown in the right-hand sub-panels). The parameters of the artificial objects are based on magnitudes and sizes measured by \citet{evstigneeva:2008} for 18 real one-component UCDs. In addition to the measured S\'{e}rsic index, we created further models with a S\'{e}rsic index in the range of $n = 1-8$. The artificial objects displayed here have an intermediate S\'{e}rsic index of $n = 4$. For each peculiar object, we selected the artificial object that is closest in magnitude. The displayed objects are sorted by decreasing magnitude ($V-$band equivalent) from panels (a) to (o). The width of a single sub-panel is 13~arcsec (1.2~kpc). In the following we list the ID of each peculiar object, as given in Tables~\ref{tab:tab1} and \ref{tab:tab3}:  (a) ID=17, (b) 15, (c) 448, (d) 838, (e) 868, (f) 711, (g) 775, (h) 867, (i) 897, (j) 869, (k) 697, (l) 5, (m) 777, (n) 735, (o) 570. The artificial objects shown in each panel are based on $r_h$ and $m_V$ of the following objects from \citet{evstigneeva:2008}: (a) UCD~1, (b) UCD~1, (c) UCD~41, (d) UCD~41, (e) UCD~41, (f) UCD~48, (g) UCD~48, (h) UCD~55, (i) UCD~55, (j) UCD~55, (k) UCD~33, (l) UCD~54, (m) UCD~21, (n) UCD~21, (o) UCD~21.}
    \label{fig:fig4}
\end{figure*}

Our visual investigation revealed peculiar compact stellar systems, which are displayed in Fig~\ref{fig:fig4}. We compare each peculiar object to an artificial compact object of similar brightness that was convolved with the PSF of the corresponding region (see Fig.~\ref{fig:fig4}, right-hand panels). The comparison illustrates that the peculiar appearance of the displayed objects is not due to PSF effects.

\subsection{Parameters and spatial distribution}
\label{sect:sect4.2}

In this section we investigate whether there is a correlation between the parameters defined in Sections~\ref{sect:sect3.2} and \ref{sect:sect3.3} and the spatial distribution of the brighter objects in our sample. For each parameter we defined a cut to separate objects with low-value from objects with high-value parameters. We then divided our sample into subsamples by combining different parameter cuts. The definition of the subsamples is specified in Table~\ref{tab:tab4}. For simplicity, we denote low core concentration with `cc', high core concentration with `CC', and analogous for residual asymmetry (ra/RA) and ellipticity (el/EL). The parameter cuts for the defined subsamples are provided in Table~\ref{tab:tab5}. We adopted the parameter cuts as well as a magnitude limit of $m_{\mathrm{Ve}} = 20.6$~mag to yield the statistically most significant differences in the distribution of cluster-centric distance between the different subsamples. We note that the cc+ra subsample contains a higher fraction of bright objects than the other subsamples, yet also has the largest spread in luminosity. There are five objects with $m_{\mathrm{Ve}} < 20.6$~mag that have no measured residual asymmetry and were therefore excluded from the following analysis.

\begin{figure*}
	\includegraphics[width=\textwidth]{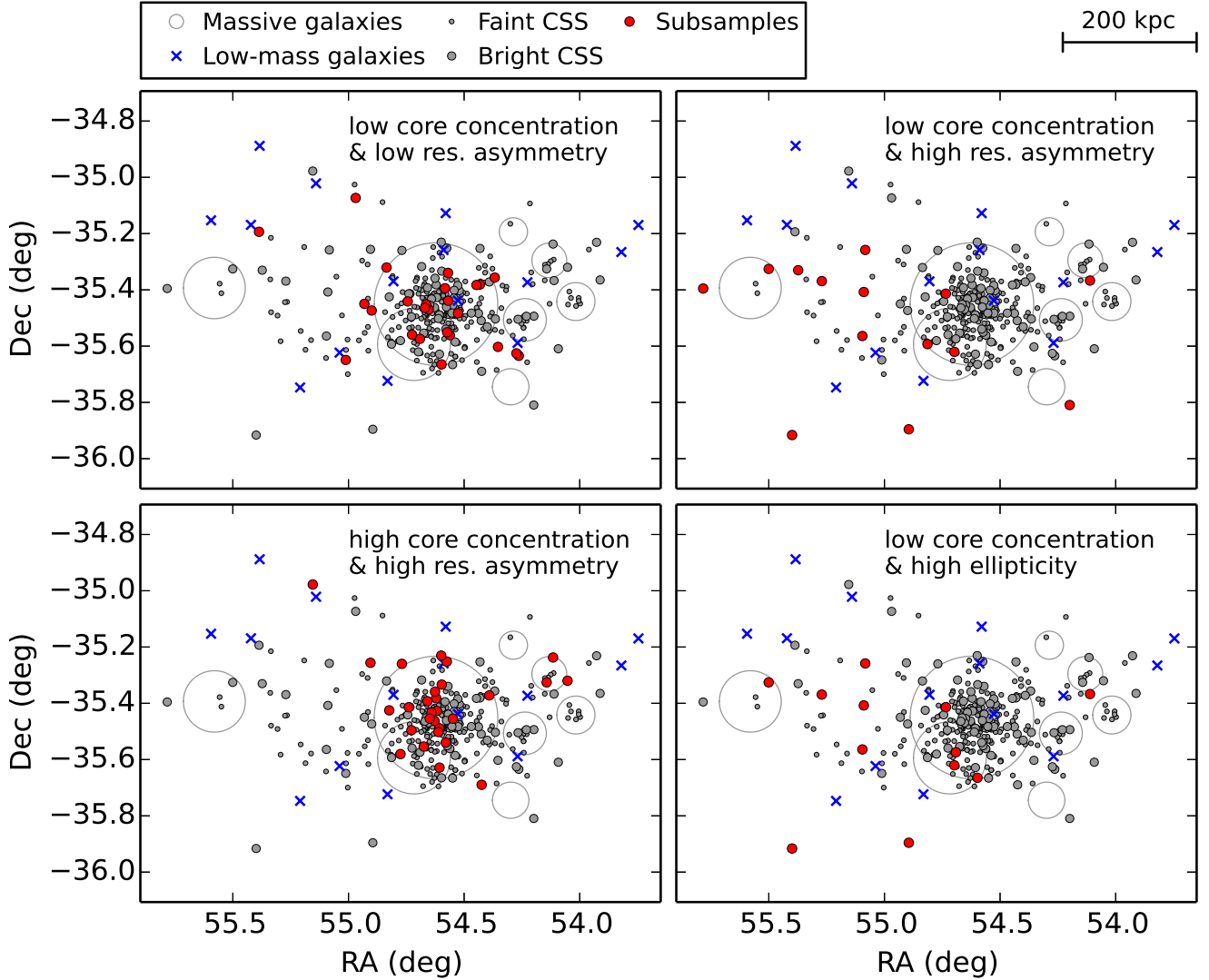}
    \caption{Spatial distribution of compact stellar systems in the Fornax cluster. For comparison we also show the distribution of the cluster galaxies. Faint CSS: compact objects with $20.6 \leq m_{\mathrm{Ve}} < 21.5$~mag. Bright CSS: compact objects with $m_{\mathrm{Ve}} < 20.6$~mag. Subsamples: cc+ra, cc+RA, CC+RA, cc+EL, as defined in Table \ref{tab:tab5}. Low-mass galaxies: galaxies with $-19 < M_r < -16$~mag from the Fornax cluster catalogue (FCC, \citealt{ferguson:1989}; based on the magnitude conversions of \citealt{weinmann:2011}). Massive galaxies: galaxies with $M_r \leq -19$~mag from the FCC. Each massive galaxy is represented by a circle with three times its isophotal diameter at $\mu_B = 25$~mag~arcsec$^{-2}$, $3\,\rm{d_{25}}$ (we used the extinction-corrected values for d$_{25}$, obtained from HyperLEDA; \citealt{makarov:2014}). The two brightest galaxies are NGC~1399 in the centre and NGC~1404 to the south-east from it.}
    \label{fig:fig5}
\end{figure*}

\begin{figure}
	\includegraphics[width=\columnwidth]{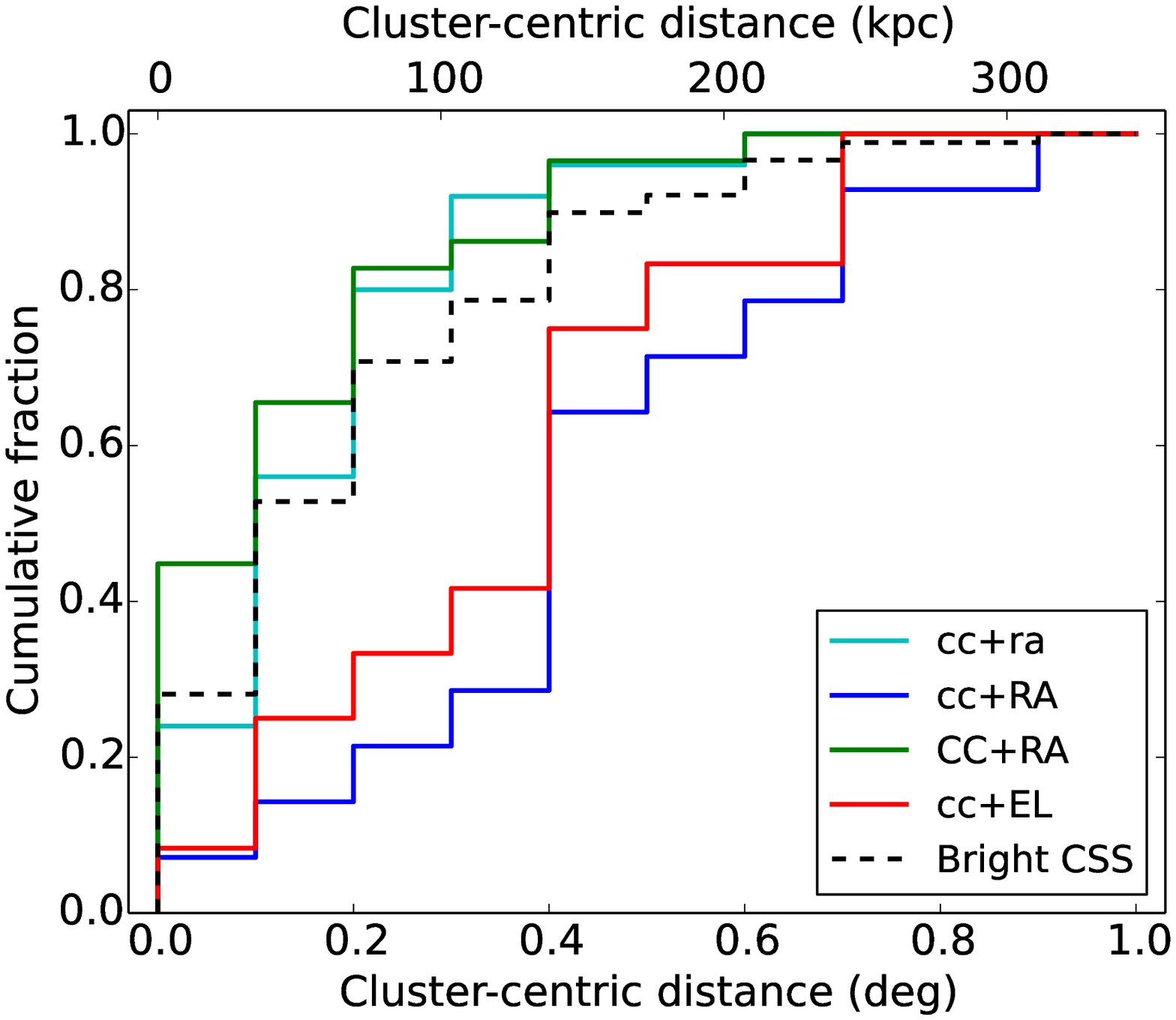}
    \caption{Cumulative distribution of projected cluster-centric distance (measured from NGC~1399) for the subsamples cc+ra, cc+RA, CC+RA and cc+EL. The distribution of all objects with $m_{\mathrm{Ve}} < 20.6$~mag (`Bright CSS') is shown for comparison.}
    \label{fig:fig6}
\end{figure}

Fig.~\ref{fig:fig5} shows the spatial distribution of our sample in the cluster, where we highlight the subsamples cc+ra, cc+RA, CC+RA and cc+EL. For comparison we also include low-mass cluster galaxies in the magnitude range $-19$~$<$~$M_r$~$<$~$-16$~mag. We probed the differences in the cluster-centric distance distributions (measured from NGC~1399) statistically with a Kolmogorov--Smirnov (KS) test, as summarized in Table~\ref{tab:tab6}, and provide a comparison of the cumulative distributions of cluster-centric distance for the defined subsamples in Fig. \ref{fig:fig6}.

We first compared the subsamples cc+ra, cc+RA, CC+RA and cc+EL to the respective other objects in the same magnitude range with $m_{\mathrm{Ve}} < 20.6$~mag. We find that both the cc+RA and the cc+EL subsamples are predominantly located at larger cluster-centric distances. This difference is statistically significant according to a KS test, with a probability of 0.0 and 0.5~per~cent, respectively, for both subsamples having the same cluster-centric distance distribution as the respective other objects in the same magnitude range. In the cc+ra and CC+RA subsamples most objects have smaller cluster-centric distances. The distribution of the cc+ra subsample is very similar to the overall distribution of respective other objects with $m_{\mathrm{Ve}} < 20.6$~mag. The distribution of the CC+RA subsample even appears to be more concentrated\footnote{This also shows that the centrally concentrated distribution of the cc+ra subsample is not related to the high fraction of bright objects in it, since the CC+RA subsample consists of fainter objects, but is even more concentrated.} than the distribution of objects in the same magnitude range, although this difference has only a low statistical significance. Compared to the low-mass cluster galaxies, we find that the cc+RA and cc+EL subsamples appear similarly distributed, whereas the distribution of the cc+ra and CC+RA subsamples seems to be more centrally concentrated.

When comparing the different subsamples to each other, we find the most significant differences between the cluster-centric distance distributions of the cc+RA subsample as compared to the cc+ra and CC+RA subsamples, with a probability of 0.0~per~cent, respectively, for having the same cluster-centric distance distribution. This is also seen for the cc+EL subsample, although with a lower statistical significance. We do not find significant differences between the distributions of the cc+ra and CC+RA subsamples. Also the distributions of the cc+RA and cc+EL subsamples are similar. But for the latter this is mainly due to the fact that the subsamples have some overlap in parameter range, since objects with high residual asymmetry can as well have high ellipticity, or vice versa.

In addition to the subsamples defined in Table~\ref{tab:tab5} we provide an alternative subsample definition in Appendix~\ref{sect:sectB}, where we set the parameter cut for the core concentration to a higher value in order to increase the number of objects in the subsamples with low core concentration. We do not find that our main results change significantly, although they have mainly lower statistical significances. For the CC+RA subsample, which includes fewer objects according to this parameter cut, we find that the more concentrated distribution, compared to the respective other objects of similar magnitude, becomes statistically more significant than previously, with a probability of 0.8~per~cent for the same distribution.

\begin{table}
 \caption{Definition of subsamples and abbreviations.}
 \label{tab:tab4}
 \begin{tabular}{ll}
  \hline
  Subsample & \\
  \hline
  cc+ra & Low core concentration and low res. asymmetry\\
  cc+RA & Low core concentration and high res. asymmetry\\
  CC+RA & High core concentration and high res. asymmetry\\
  cc+EL & Low core concentration and high ellipticity\\
  \hline
 \end{tabular}
\end{table}

\begin{table*}
 \caption{Parameter ranges for the subsamples cc+ra, cc+RA, CC+RA and cc+EL, defined such that they yield the statistically most significant differences in the distribution of cluster-centric distance. For each subsample we give the fraction of objects in the respective magnitude range and the overlap fractions with the artificial objects.}
 \label{tab:tab5}
 \begin{tabular}{lllllll}
  \hline
  Subsample & Cuts for $m_{\mathrm{Ve}} < 20.0$~mag & Objects & Art. objects & Cuts for $20.0 \leq m_{\mathrm{Ve}} < 20.6$~mag & Objects & Art. objects\\
  && (per cent) & (per cent) & & (per cent) & (per cent)\\
  \hline
  cc+ra & cc < 0.77 and ra < 0.5 & 46.9 & 21.2 & cc < 0.83 and ra < 0.55 & 17.5 & \phantom{0}0.0\\
  cc+RA & cc < 0.77 and ra > 0.5 & \phantom{0}9.4 & \phantom{0}0.2 & cc < 0.83 and ra > 0.55 & 19.3 & \phantom{0}0.0\\
  CC+RA & cc > 0.77 and ra > 0.5 & 25.0 & \phantom{0}5.6 & cc > 0.83 and ra > 0.55 & 36.8 & \phantom{0}3.7\\
  cc+EL & cc < 0.77 and el\, > 0.2 & 12.5 & \phantom{0}0.2 & cc < 0.83 and el\, > 0.22 & 14.0 & \phantom{0}0.0\\
  \hline
 \end{tabular}
\end{table*}

\begin{table}
 \caption{KS test probabilities (percentage) for the null hypothesis that two subsamples have the same cluster-centric distance distribution. In the last row the distributions of the individual subsamples are compared to the respective other compact objects with different parameters in the same magnitude range with $m_{\mathrm{Ve}} < 20.6$~mag.}
  \label{tab:tab6}
 \begin{tabular}{lllll}
  \hline
  Subsample & cc+ra & cc+RA & CC+RA & cc+EL\\
  \hline
  cc+ra & 100.0 & \phantom{00}0.0 & \phantom{0}55.3 & \phantom{00}0.4\\
  cc+RA & & 100.0 & \phantom{00}0.0 \phantom{0} & \phantom{0}98.7\\
  CC+RA & & & 100.0 & \phantom{00}1.0\\
  cc+EL & & & & 100.0\\
  respective & \phantom{0}41.7 & \phantom{00}0.0 & \phantom{00}6.7 \phantom{0} & \phantom{00}0.5\\
  other CSS &&&&\\
  \hline
 \end{tabular}
\end{table}

\subsection{Distribution in phase-space}
\label{sect:sect4.3}

\begin{figure*}
	\includegraphics[width=\textwidth]{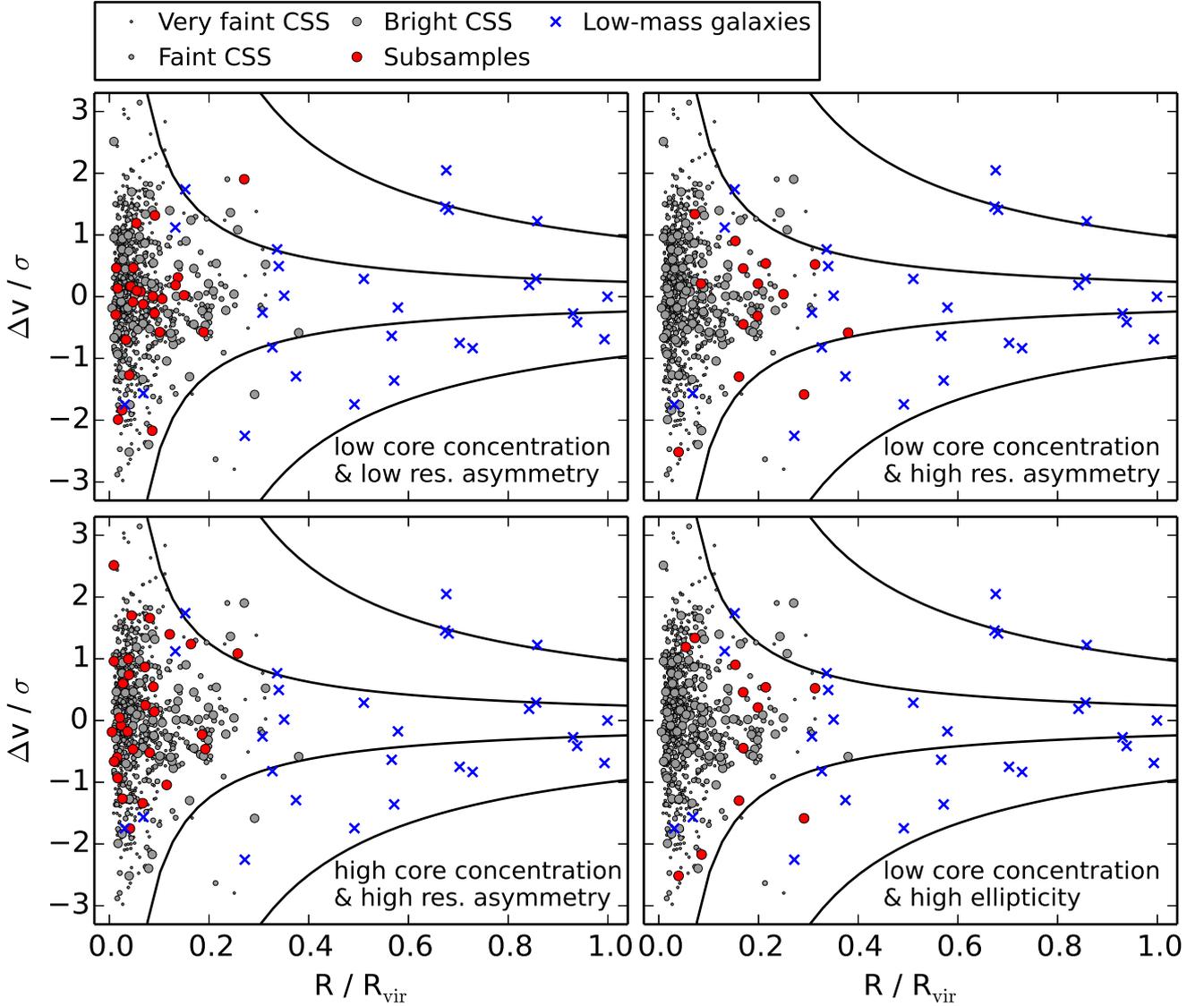}
    \caption{Phase-space distribution of compact stellar systems and low-mass galaxies in the Fornax cluster. Very faint CSS: compact objects with $m_{\mathrm{Ve}} \geq 21.5$~mag (not part of our working sample). Faint CSS: compact objects with $20.6 \leq m_{\mathrm{Ve}} < 21.5$~mag. Bright CSS: compact objects with $m_{\mathrm{Ve}} < 20.6$~mag. Subsamples: cc+ra, cc+RA, CC+RA, cc+EL, as defined in Table \ref{tab:tab5}. Low-mass galaxies: galaxies with $-19 < M_r < -16$~mag from the FCC (\citealt{ferguson:1989}; based on the magnitude conversions of \citealt{weinmann:2011}). $\Delta v$ is the relative velocity of an object with respect to the cluster mean velocity (1460~km~s$^{-1}$). We denote $\sigma$ as the cluster velocity dispersion (324~km~s$^{-1}$), $R$ as the cluster-centric distance, and $R_{\mathrm{vir}}$ as its virial radius ($2.5\degr$, see Section~\ref{sect:sect4.3}). The mean velocity and dispersion were calculated from all compact stellar systems and FCC galaxies within a cluster-centric distance of 1.0\degr. The solid lines correspond to caustic lines of constant $(\Delta v / \sigma)~\times~(R/R_{\mathrm{vir}})$ at $\pm$0.1 and $\pm$0.4, respectively.}
    \label{fig:fig7}
\end{figure*}

In order to examine the phase-space distribution of compact stellar systems in the Fornax cluster, we define $\Delta v$ as an object's relative velocity with respect to the cluster mean velocity (1460~km~s$^{-1}$). The latter is the average velocity of Fornax cluster members within a cluster-centric distance of 1\degr, including all compact objects from our catalogue and galaxies from the Fornax cluster catalogue (FCC, \citealt{ferguson:1989}). We denote the corresponding standard deviation (324~km~s$^{-1}$) as the cluster velocity dispersion $\sigma$, $R$ as the cluster-centric distance, and $R_{\mathrm{vir}}$ as its virial radius (assumed to be 0.85~Mpc or 2.5\degr, which is the average of the 0.7~Mpc quoted by \citealt{drinkwater:2001} and 1.0~Mpc quoted by \citealt{murakami:2011}). The phase-space distribution is shown in Fig.~\ref{fig:fig7} as $\Delta v / \sigma$ versus $R/R_{\mathrm{vir}}$. We highlight the subsamples cc+ra, cc+RA, CC+RA and cc+EL, and further discriminate between very faint compact stellar systems in the magnitude range of GCs ($m_{\mathrm{Ve}} \ge 21.5$~mag, which are not part of our working sample), faint objects ($20.6 \leq m_{\mathrm{Ve}} < 21.5$~mag), and bright objects ($m_{\mathrm{Ve}} < 20.6$~mag), and include low-mass galaxies in the magnitude range $-19 < M_r < -16$~mag for comparison. Table~\ref{tab:tab7} summarizes the velocity dispersion for the various subsamples and cluster populations. We show the phase-space distribution of the alternative subsamples in Appendix~\ref{sect:sectB} (Fig.~\ref{fig:fig10} and Table~\ref{tab:tab10}).

Phase-space diagrams allow one to study the accretion history of a cluster population. For example, \citet{noble:2013} used caustic profiles, which are lines of constant $(\Delta v / \sigma)~\times~(R/R_{\mathrm{vir}})$, to distinguish between infalling and virialized cluster members, where higher values trace systems that were more recently accreted. In Fig.~\ref{fig:fig7} we plotted caustic lines of constant $(\Delta v / \sigma)~\times~(R/R_{\mathrm{vir}})$ at $\pm$0.1 and $\pm$0.4, respectively. Cosmological simulations show that early accreted systems would be predominantly located within the inner caustic lines of $\pm$0.1, systems accreted later between the caustics of $\pm$0.1 and $\pm$0.4, and recently accreted systems along or outside the caustic lines of $\pm$0.4 \citep[cf.][]{haines:2012, noble:2013}.

We find that essentially all compact stellar systems are located within the inner caustics of $\pm$0.1. Also some low-mass galaxies are found in this region, but many of them occupy the region in between the caustic lines of $\pm$0.1 and $\pm$0.4. This is also reflected in the velocity dispersion, where we find the lowest dispersion in particular for the compact stellar systems with $m_{\mathrm{Ve}} < 21.5$~mag (the samples comprising the bright and faint CSS) and the highest for the low-mass galaxies.

Among the bright compact stellar systems with $m_{\mathrm{Ve}}~<~20.6$~mag, we do not find statistically significant differences in the velocity distribution of the four different subsamples. The main difference seems to be in the spatial distribution. However, with respect to the velocity dispersion, we find that the cc+ra subsample has on average the lowest and the cc+EL subsample the highest velocity dispersion. The two subsamples cc+RA and CC+RA with large residual asymmetry have velocity dispersions in between. In the phase-space diagram we note that especially the cc+EL subsample is predominantly distributed along the inner caustic lines.

\begin{table}
 \caption{Velocity dispersion for very faint ($m_{\mathrm{Ve}} \geq 21.5$~mag, not part of our working sample), faint ($20.6 \leq m_{\mathrm{Ve}} < 21.5$~mag) and bright ($m_{\mathrm{Ve}} < 20.6$~mag) compact stellar systems, low-mass galaxies (FCC), and the subsamples cc+ra, cc+RA, CC+RA, cc+EL, as defined in Table~\ref{tab:tab5}. The velocity dispersion of each subsample is calculated as standard deviation within a cluster-centric distance of $R \leq 1.0\degr$  ($\sigma_{\mathrm{tot}}$), $R \leq 0.4\degr$ ($\sigma_{\mathrm{in}}$) or $0.4 < R \leq 1.0\degr$ ($\sigma_{\mathrm{out}}$), based on the velocities given in Table~\ref{tab:tab1}. $N_{\mathrm{obj}}$ corresponds to the number of objects from the respective subsample in the inner ($N_{\mathrm{obj,in}}$) or outer ($N_{\mathrm{obj,out}}$) cluster region. The velocity dispersion is given in km~s$^{-1}$. $1.0\degr$ corresponds to $0.346$ and $0.4\degr$ to $0.138$~Mpc at the distance of Fornax.}
 \label{tab:tab7}
 \begin{tabular}{llllll}
  \hline
  Subsample & $\sigma_{\mathrm{tot}}$ & $\sigma_{\mathrm{in}}$ & $N_{\mathrm{obj,in}}$& $\sigma_{\mathrm{out}}$ & $N_{\mathrm{obj,out}}$\\
  \hline
  Very faint CSS & 337 & 336 & 535 & 341 & 14\\
  Faint CSS & 297 & 302 & 232 & 252 & 29\\
  Bright CSS & 307 & 316 & \phantom{0}73 & 271 & 21\\
  Low-mass & 400 & 505 & \phantom{00}4 & 317  & \phantom{0}7\\
  galaxies&&&&&\\
  cc+ra & 303 & 282 & \phantom{0}23 & 401 & \phantom{0}2\\
  cc+RA & 326 & 485 & \phantom{00}4 & 330 & 10\\
  CC+RA & 328 & 338 & \phantom{0}25 & 246 & \phantom{0}4\\
  cc+EL & 414 & 556 & \phantom{00}5 & 269 & \phantom{0}7\\
  \hline
 \end{tabular}
\end{table}

\section{Discussion}
\label{sect:sect5}
\subsection{Analysis methods and limitations}
\label{sect:sect5.1}

The seeing FWHM varies over the mosaic, since the latter is based on images acquired in multiple nights and observing runs. In the PSF analysis we tried to account for this by determining the PSF in regions of similar FWHM, as indicated in Fig.~\ref{fig:fig2}. In {\it Region 3} of our mosaic we observe systematically lower core concentration for our sample with $m_{\mathrm{Ve}} < 21.5$~mag. We investigated whether this effect could be due to a non-matching PSF. We therefore fitted PSFs from other regions on the mosaic to all objects from our sample that are located in {\it Region 3}. We find that the core concentration decreases when a PSF of smaller FWHM is used. Thus, if there are variations of the PSF FWHM on scales smaller than the considered region within which the PSF is determined, an artificially low core concentration can occur if an object is located in an area with locally larger FWHM than the PSF FWHM of that region.

In order to estimate local variations of the FWHM we computed the minimum FWHM around each object. To find the minimum FWHM we first measured the FWHM of all objects in {\it Region 3} with \texttt{SExtractor}. From all \texttt{SExtractor}-detected objects brighter than $m_{\mathrm{Ve}} = 22$~mag we then determined the minimum FWHM within a circular area of 450~pixels (1.8~arcmin) in radius around each object. Finally we compared the local FWHM to the FWHM of the PSF stars from this region. In {\it Region 3} we neither find a relation of core concentration with local FWHM, nor are all objects with low core concentration systematically located in areas with large local FWHM.

We extended this test to the full sample of compact stellar systems brighter than $m_{\mathrm{Ve}} = 20.6$~mag and investigated how much a locally varying FWHM would affect our results from Section~\ref{sect:sect4.2}. We first computed the relative local FWHM for each object as difference between the measured local FWHM and the average FWHM of all PSF stars of the respective region. Since we found a slight dependence of the core concentration on the relative local FWHM for objects brighter than $m_{\mathrm{Ve}} = 20.6$~mag, we defined a linearly varying parameter cut that follows this relation. We then defined new subsamples, using the parameter cuts for the residual asymmetry and ellipticity as specified in Table~\ref{tab:tab5}, but with a linearly varying parameter cut for the core concentration. According to this definition, our main result that the subsamples with low core concentration and high residual asymmetry (or high ellipticity) are mainly distributed at larger cluster-centric distances remains unchanged\footnote{With probabilities of 0.0 and 0.8~per~cent, respectively, for the two subsamples having the same cluster-centric distance distribution as objects in the same magnitude range according to a KS test.}.\\

The angular resolution of our data limits our ability to identify possible blends of close neighbouring sources. \citet{richtler:2005} resolved one of the peculiar compact objects (Fig.~\ref{fig:fig4}, panel g) into two sources. We therefore attempt to estimate how likely it is that all objects with significant asymmetry or ellipticity are blends. For this purpose we assume that objects that appear asymmetric or elongated could be blends if a close, up to three magnitudes fainter, neighbour source was located within a distance of $4 - 7$ pixels (corresponding to 91 and 160~pc, respectively). When considering all objects from our bright sample ($m_{\mathrm{Ve}} < 20.6$~mag) with high residual asymmetry or high ellipticity (according to the definition in Table~\ref{tab:tab5}), which have cluster-centric distances between 20 and 160~kpc, we find an expected number density of faint sources of 0.09 arcsec$^{-2}$ if all those objects were blends. For comparison, the observed number density of faint sources with magnitudes $20.6 < m_{\mathrm{Ve}} < 23.6$~mag located within the same area is 0.003 objects arcsec$^{-2}$. Thus, only a single one of the 32 bright asymmetric or elongated objects we observe in this area would statistically be expected to be a blend. We note, however, that an overdensity of GCs within 1~kpc of brighter compact stellar systems, including a fraction within 300~pc, was recently reported by \citet{voggel:2016}. We may thus expect a few blends among the said objects.

\subsection{Peculiar compact stellar systems in the Fornax cluster}
\label{sect:sect5.2}

We report the discovery of peculiar compact stellar systems in the Fornax cluster, which appear asymmetric or elongated in our images (illustrated in Fig.~\ref{fig:fig4}). The presence of a few peculiar objects in Fornax was already noted by \citet{richtler:2005} and \citet{voggel:2016}\footnote{Two of our displayed objects were shown in \citet{richtler:2005} (Fig.~\ref{fig:fig4}, panels g and n). The object in panel (n) was also pointed out by \citet{voggel:2016} to be peculiar.}.

Some compact Fornax cluster members have previously been observed with \textit{HST} \citep{evstigneeva:2008}. Two of them are also shown in Fig.~\ref{fig:fig4} (panels d and h), but we note that a direct comparison of the inner structure is not possible due to our much broader PSF. At the same time the \textit{HST} images are too shallow for a comparison of the outer, low-surface-brightness structure that we are able to measure in our data. The \emph{azimuthally averaged} UCD surface-brightness profiles of \citet{evstigneeva:2008} reach 26~mag~arcsec$^{-2}$ at best (see their fig.~1), whereas we reach an image depth of 26.8~mag~arcsec$^{-2}$ at $S/N=1$ \emph{per pixel} (see Section~\ref{sect:sect2.1}).

In the Virgo cluster the structure of a large sample of UCDs has been investigated by \citet{liu:2015}, revealing faint envelopes around many of the objects. However, objects with similarly asymmetric appearance as observed in our data were not reported. In the cluster Abell S0740, \citet{blakeslee:2009} detected faint envelopes around candidate UCDs, which show signs of being elongated in \textit{HST} images.

There are two main formation scenarios discussed in the literature, relating UCDs to either the population of galaxies or to the population of star clusters. In the former case, it is suggested that UCDs may be the remnant nuclei of tidally stripped nucleated dwarf galaxies \citep[e.g.][]{bekki:2003, pfeffer:2013}. In a star cluster origin, UCDs may grow to sizes and masses larger than typical GCs, if they form in star cluster complexes by merging of young massive star clusters \citep[e.g.][]{fellhauer:2005}.

According to simulations by \citet{bekki:2003}, observable signatures of a tidal stripping origin would be tidal tails as well as relics from the envelope of the progenitor galaxy that has been stripped. \citet{lisker:2016} did not find any diffuse streams that would be signs of tidal debris around any of the 904 spectroscopically confirmed objects in our compiled catalogue. As noted by \citet{lisker:2016}, such streams should be visible in most parts of the mosaic if their surface brightness level is in the range $27.5 \lesssim \mu_{\mathrm{Ve}} \lesssim 28.0$~mag~arcsec$^{-2}$. The absence of visible tidal debris does not necessarily need to be in contradiction to a tidal stripping origin of these objects. \citet{pfeffer:2014} showed, based on cosmological simulations, that most low-mass cluster galaxies were disrupted already many gigayears ago. Thus, if most of the compact stellar systems resulted from early stripping events, we simply might not be able to observe relics of tidal debris any more today, since tidal tails disperse on time-scales of a few Gyr \citep{pfeffer:2013}. However, \citet{bruens:2012} predicted that an object formed via merging of massive star clusters in a star cluster complex would also be surrounded by a faint stellar envelope, thus exhibiting a similar two-component surface-brightness profile as reported in the tidal stripping simulations. Therefore, without the detection of the predicted long tidal streams, we cannot discriminate between a stripping or a merging origin.

In the star cluster merging scenario described by \citet{bruens:2011}, the forming object can look quite asymmetric as long as the merger is not yet complete. However, the authors showed that star clusters in a cluster complex typically merge on time-scales of only a few hundred Myr. As a consequence, if the structures we observe for the peculiar systems in Fig.~\ref{fig:fig4} are signatures of an ongoing star cluster merger, then these should be comparatively young systems. This seems to be in contradiction with the observation that UCDs have in general old stellar populations \citep{evstigneeva:2007, paudel:2010, francis:2012}.

On the other hand, \citet{fellhauer:2005} reported that a merger remnant would be stable over a time-scale of 10~Gyr. Thus, compact objects may have formed from star cluster merging already in very early phases during the formation of the Fornax cluster core, when the merging of gas-rich galaxies possibly offered conditions for strong starbursts in which large star cluster complexes are thought to form. At the same time \citet{fellhauer:2005} demonstrated that such a merger remnant constantly loses some of its mass with every pericentric passage due to the tidal field of the cluster. Therefore the peculiar structures of some compact systems may not stem from an ongoing merger of star clusters, but may be associated with the deformation or disruption of an extended star cluster that formed via star cluster merging when the Fornax cluster core assembled. Disruption signatures can thus not necessarily discriminate between the proposed formation scenarios of compact stellar systems.

To estimate whether tidal stripping would be efficient for disturbing the outer structure of an UCD-like compact object orbiting in the Fornax cluster, we calculate the tidal radius according to \citet{king:1962}:
\begin{equation}
R_{\mathrm{tidal}} = R_{\mathrm{peri}} \left( \frac{M_{\mathrm{obj}}}{M_{\mathrm{cl}}(R_{\mathrm{peri}})\,(3 + e)}\right)^{1/3},
	\label{eq:eq4}
\end{equation}
where $R_{\mathrm{peri}}$ is the pericentric distance, $M_{\mathrm{obj}}$ the total mass of the object, $M_{\mathrm{cl}}(R_{\mathrm{peri}})$ the enclosed cluster mass within $R_{\mathrm{peri}}$ and $e$ the eccentricity of the orbit for which we adopt a value of 0.5\footnote{Bound orbits have an eccentricity of $e < 1.0$. An eccentricity of $e = 0.0$ would correspond to a circular orbit, $e > 0.7$ to a highly eccentric orbit.}. For a typical UCD-like object with a mass of $M_{\mathrm{obj}} = 10^7$~M$_{\sun}$, when assuming it reaches an orbital pericentre of $R_{\mathrm{peri}} = 20$~kpc, the tidal radius would be in the range of $200 - 300$~pc, depending on the adopted mass profile for the Fornax cluster \citep{drinkwater:2001, richtler:2008, SRH2010}. While this is already on the order of 10 effective radii for a UCD of mass $M_{\mathrm{obj}} = 10^7$~M$_{\sun}$ \citep[cf.][]{norris:2014}, we would be able to observe such an object out to its tidal radius in our deep imaging data, according to the surface-brightness profiles of typical UCDs \citep[see fig.~1 of][]{depropris:2005}. This estimate shows that distortions of the outer structure due to tidal stripping could be expected for objects with close cluster-centric passages. However, since the tidal radius of an object solely depends on its total mass, and not on how the object mass is distributed, it is not possible to infer the nature of the disturbed object, whether it is the remains of a stripped galaxy or an extended star cluster in process of disruption.

For the above estimate of the tidal radius we assumed that the objects would be only influenced by the cluster's tidal field. None the less, some objects may be more strongly affected by very close bright galaxies, especially further out from the cluster centre. \citet{schuberth:2008} investigated whether some compact stellar systems with $M_V < -9.5$~mag were consistent with being associated with any of the ten brightest galaxies in the Fornax cluster core, according to their spatial and velocity distribution. The authors showed that among the compact objects, which are located within a projected distance of $1.5\,\rm{d_{25}}$\footnote{d$_{25}$ is the isophotal diameter of a galaxy at $\mu_B$~$=$~$25$~mag~arcsec$^{-2}$.} from a bright galaxy, the metal-rich (red) objects have velocities not deviating by more than 100~km~s$^{-1}$ from the velocity of the closest massive galaxy, whereas the metal-poor (blue) objects seem to be characterized by a larger spread in velocities. The remaining objects at distances larger than $1.5\,\rm{d_{25}}$ from any bright galaxy seem to be consistent with being kinematically associated with the extended GC system of NGC~1399, out to a cluster-centric distance of 30~arcmin.

In Fig.~\ref{fig:fig5}, showing the spatial distribution of known Fornax cluster members, we represented each massive galaxy with $M_r \leq -19$~mag by three times its isophotal diameter ($3\,\rm{d_{25}}$). We find that only few objects from our subsamples lie close in projection to a bright galaxy other than NGC~1399, and of these only a minor fraction also has a similar velocity\footnote{We find that the following number of objects from our subsamples are located within a distance of  $\rm{r} = 1.5\,\rm{d_{25}}$ from a massive galaxy with $M_r \leq -19$~mag other than NGC~1399: 4 out of 25 in cc+ra, 3 out of 14 in cc+RA, 7 out of 29 in CC+RA, and 4 out of 12 in cc+EL. When we require additionally a relative velocity smaller than $\Delta v = 200$~km~s$^{-1}$, the number of objects is: 1 in cc+ra, 1 in cc+RA, 3 in CC+RA, and 2 in cc+EL.}. Therefore we see no indication that a significant fraction of objects from our subsamples is bound to massive galaxies with $M_r \leq -19$~mag other than NGC~1399.

\subsection{Spatial and phase-space distribution of compact stellar systems}
\label{sect:sect5.3}

In general it is observed that GCs and galaxies have different spatial distributions in galaxy clusters. For example \citet{zhang:2015} observed that in the central core region of Virgo, the dwarf ellipticals (dEs) have a much flatter number density profile than the GCs. In Fornax it was also found that the GC population of NGC~1399 is much more centrally concentrated than the surrounding dEs \citep{GDE2009, hilker:2011}. \citet{GDE2009} reported that the UCDs in Fornax have lower velocity dispersions than both the GCs and the low-mass galaxies in the core region, which we also find in this work. \citet{SRH2010} further distinguished between red and blue GCs, and their analysis showed that UCDs have a velocity dispersion lower than that of the blue GCs, but higher than that of the red GCs.

In Sections~\ref{sect:sect4.2} and \ref{sect:sect4.3} we investigated the spatial and phase-space distribution of compact stellar systems in the Fornax cluster, where we focused on the distribution of objects with $m_{\mathrm{Ve}} < 20.6$~mag ($M_{\mathrm{Ve}} < -11.1$~mag), corresponding to the brighter UCD magnitude range. We compared the differences in the cluster-centric distance distributions of smaller subsamples, where we categorized the objects according their core concentration, residual asymmetry, and ellipticity. In the following we discuss the results on the spatial and phase-space distribution of our subsamples in terms of the origin and nature of compact stellar systems.

We find that the objects in the subsample with low core concentration and high residual asymmetry, as well as in the subsample with low core concentration and high ellipticity, are predominantly distributed at larger cluster-centric distances, compared to objects in the same magnitude range but with different parameters\footnote{We note that the subsamples with larger cluster-centric distances may even be slightly under-represented in our study, as compared to the other subsamples, if the completeness varies with cluster-centric distance (see Section~\ref{sect:sect2.4}).}. Their extended galaxy-like spatial distribution might favour a formation scenario in which they are remnants of stripped low-mass galaxies\footnote{This does not exclude that some of them formed as massive star clusters in the tidal tails of a major galaxy merger \citep[e.g.,][]{gallagher:2001}.}. However, since their location at larger cluster-centric distances implies a currently large tidal radius, their structure could only be explained by tidal stripping if these objects are on rather eccentric orbits with small pericentre distances, such that they possibly approached the cluster centre very closely at earlier times. We observe that especially the subsample with low core concentration and high ellipticity has high velocities relative to the Fornax cluster, which would be expected for radial orbits. This subsample is mainly distributed \emph{along} the inner caustic lines of constant $(\Delta v / \sigma)~\times~(R/R_{\mathrm{vir}}) = \pm0.1$ and seems to integrate smoothly into the phase-space distribution of low-mass cluster galaxies. This could indicate that these objects, or their progenitors, may have been accreted more recently compared to the overall population of compact stellar systems in the UCD and GC magnitude range, which are predominantly confined \emph{within} those caustics.

In the Virgo cluster, observations of a large sample of UCDs around M87 point in a similar direction: UCDs at cluster-centric distances larger than 40~kpc have a radially biased orbital structure consistent with that of stripped galaxies on radial orbits \citep{zhang:2015}. Although no compact stellar systems with obvious distorted outer structures have been reported in Virgo so far, \citet{liu:2015} observed that compact objects with signs of a faint stellar envelope are found only at distances larger than $0.1\degr$   (corresponding to 28,8~kpc at a distance of 16.5~Mpc for Virgo; \citealt{blakeslee:2009}) from the two brightest cluster galaxies M87 and M49, respectively. Their distribution is, however, still more concentrated than that of the cluster's dwarf galaxy population. The authors interpret this sequence of decreasing envelope fraction with decreasing cluster-centric distance as an indication that tidal stripping of nucleated dwarf galaxies plays an important role for the population of compact stellar systems in Virgo.

In Fornax, when defining a subsample of objects with no signs of distortions\footnote{By choosing objects with low residual asymmetry and low ellipticity according to Table~\ref{tab:tab5}, with parameter cuts at ra < 0.5 and el < 0.2 for objects with $m_V < 20.0$~mag, and ra < 0.55 and el < 0.22 for objects with $20.0 \leq m_V < 20.6$~mag.}, we find the majority of these objects distributed at small cluster-centric distances. This would at least not imply \emph{recent} stripping, if these objects were stripped galaxies at all. Instead, these objects might already be stripped down to their tidal radius. This could be expected if the objects had been accreted at very early times, and had been exposed for longer to the cluster tidal field than, e.g., the objects with low core concentration and high residual asymmetry (or high ellipticity). This would be supported by their lower velocity dispersion, compared to other subsamples. \citet{smith:2015}, who studied the effects of harassment on early-type dwarf galaxies in galaxy clusters with numerical simulations, concluded that strongly harassed galaxies would be characterized by low orbital velocities.

However, for this subset of objects with no signs of distortions, we would not exclude a star cluster origin either. Based on the spatial and phase-space distribution alone, it may be difficult to disentangle a star cluster from a galaxy origin --- possible progenitor galaxies may have been accreted very early on, or even formed at the same epoch as massive star clusters. For example, \citet{bournaud:2008} predicted that during the formation of giant ellipticals through mergers of gas-rich galaxies, tidal dwarf galaxies ($M_* = 10^8 - 10^9$~M$_{\sun}$) could form along with massive star clusters ($M_* = 10^5 - 10^7$~M$_{\sun}$).

Notwithstanding these considerations, the best candidates for a star cluster origin among the considered subsamples seem to be the objects with high core concentration and high residual asymmetry. Their distribution is very centrally concentrated and looks quite different from the distribution one would expect for galaxies. Moreover, this subsample is characterized by a higher core concentration, i.e. having a more star-like central component. Possibly we observe here a population of bright deformed star clusters instead of remnants of stripped galaxies. Since most of the objects are in close proximity to NGC~1399 the tidal radius would be quite small so that current stripping or deformation could indeed be expected. Some objects of this subsample are located in between NGC~1399 and the close elliptical galaxy NGC~1404 to the south-east from it (see Fig.~\ref{fig:fig5}). \citet{kim:2013} and \citet{dabrusco:2016} took an overabundance of GCs between NGC~1399 and NGC~1404 (as well as other surrounding galaxies) as an indication for interactions in the recent past, which was also simulated by \citet{bekki:2003b}. Thus, some of our objects with high core concentration and high residual asymmetry may have been distorted and also freed from their parent galaxies during possible interactions of massive cluster galaxies in the Fornax cluster core.

\section{Conclusions}
\label{sect:sect6}
Our deep imaging data of the Fornax cluster core reveal peculiar compact stellar systems, which appear asymmetric or elongated. From their structure alone we cannot infer their origin,  whether these objects are luminous star clusters in the process of disruption or possible remnants of stripped galaxies.  However, the spatial distribution of objects with low core concentration and high residual asymmetry (or high ellipticity) at mainly larger cluster-centric distances may be explained with a galaxy origin of at least some of these objects. This is also supported by the high relative velocities we observe in particular for the objects with low core concentration and high ellipticity.

\section*{Acknowledgements}
This paper is based on observations collected at the European Organisation for Astronomical Research in the Southern Hemisphere, Chile (programmes 082.A-9016 and 084.A-9014).
CW\ is a member of the International Max Planck Research School for Astronomy and Cosmic Physics at the University of Heidelberg (IMPRS-HD).
CW\ acknowledges support by ESO for a research visit while this work was in progress.
TL\ acknowledges support by ESO through the Scientific Visitor Programme while this work was in progress.
TL\ was supported within the framework of the Excellence Initiative by the German Research Foundation (DFG) through the Heidelberg Graduate School of Fundamental Physics (grant number GSC 129/1) and through the Innovation Fund FRONTIER of Heidelberg University.
TL thanks Michael Drinkwater for having shared a Fornax cluster object catalogue while this project was at a preliminary stage.
We thank the anonymous referee for useful comments that helped improving the presentation of our results.
This research has made use of the SIMBAD data-base, operated at CDS, Strasbourg, France.
This research has made use of the VizieR catalogue access tool, CDS, Strasbourg, France. The original description of the VizieR service was published in A\&AS 143, 23.
This research has made use of the HyperLeda data-base (http://leda.univ-lyon1.fr).





\bibliographystyle{mnras}
\bibliography{mybibliography} 





\section*{Supporting information}

Additional Supporting Information can be found in the online version
of this article:\\

\noindent
\textbf{Table 1.} Compilation of compact stellar systems known to be Fornax cluster members.

\noindent
\textbf{Table 3.} Parameter catalogue for the working sample of spectroscopically confirmed Fornax cluster members with $m_{\mathrm{Ve}} < 21.5$~mag.

\noindent
(http://www.mnras.oxfordjournals.org/lookup/suppl/
doi:10.1093/mnras/stw827/-/DC1).\\

\noindent
Please note: Oxford University Press is not responsible for the content or functionality of any supporting materials supplied by the authors. Any queries (other than missing material) should be directed to the corresponding author for the article.


\appendix
\section{Parameter relations}
\label{sect:sectA}

\begin{figure}
	\includegraphics[width=\columnwidth]{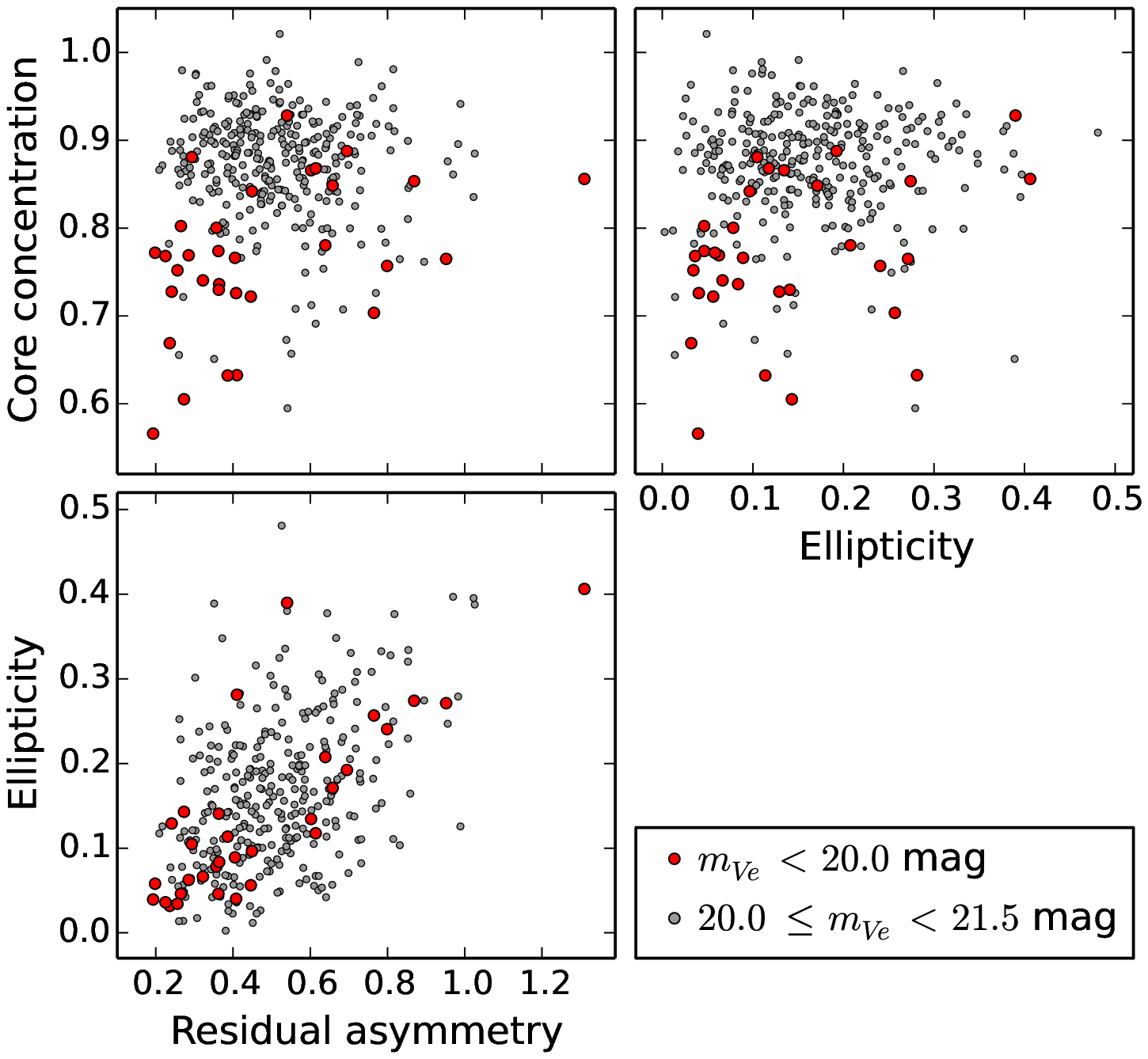}
    \caption{Parameter relations for compact stellar systems from our working sample.}
    \label{fig:fig8}
\end{figure}

In Fig.~\ref{fig:fig8} we display the relations between the parameters core concentration, residual asymmetry, and ellipticity, which we derived in Section~\ref{sect:sect3} for our working sample.

\section{Alternative subsample definition}
\label{sect:sectB}
\begin{figure*}
	\includegraphics[width=0.79\textwidth]{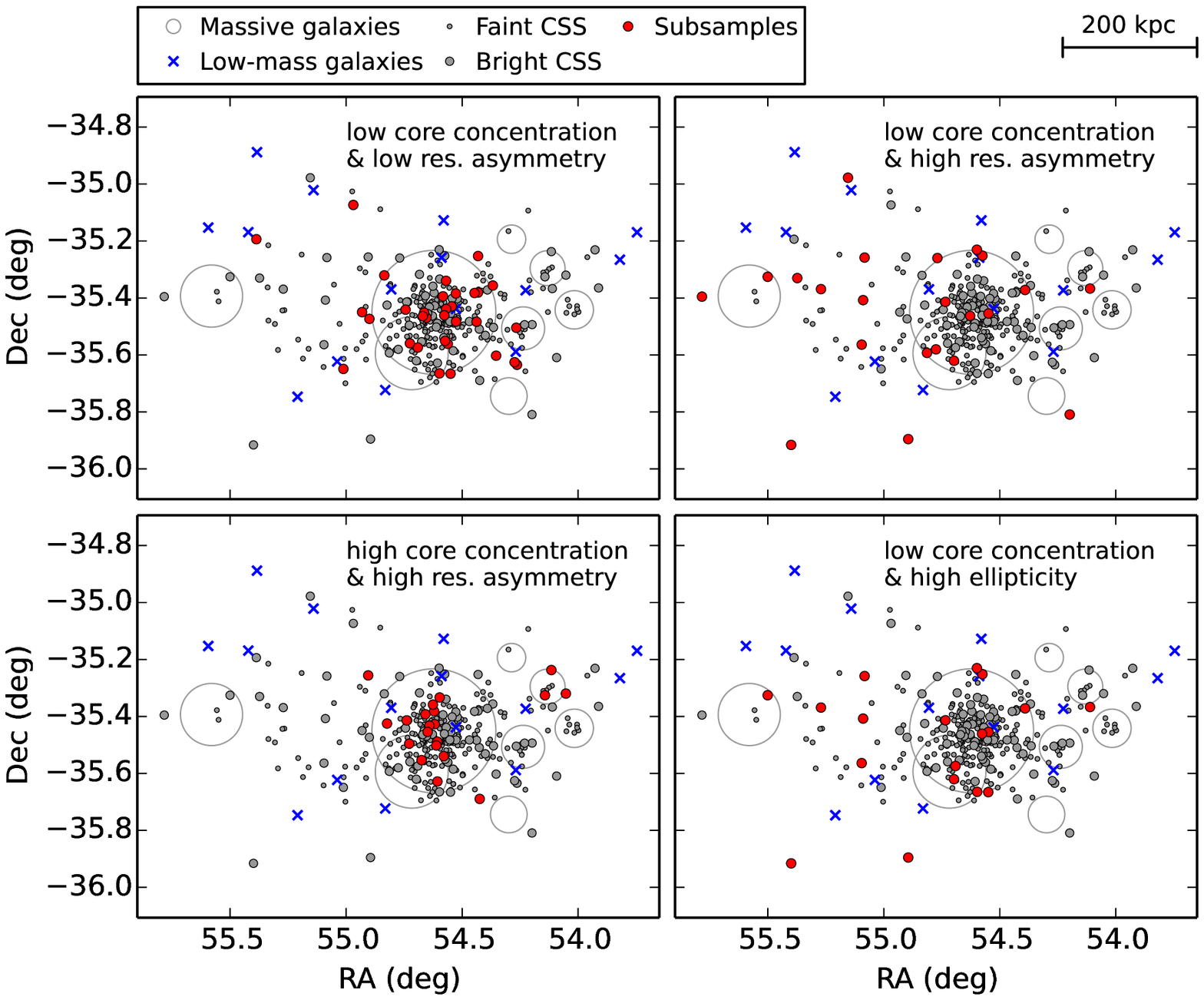}
    \caption{Spatial distribution of compact stellar systems in the Fornax cluster. Same as Fig.~\ref{fig:fig5}, but illustrating the alternative subsamples. For comparison we also show the distribution of the cluster galaxies. Faint CSS: compact objects with $20.6 \leq m_{\mathrm{Ve}} < 21.5$~mag. Bright CSS: compact objects with $m_{\mathrm{Ve}} < 20.6$~mag. Alternative subsamples: cc+ra, cc+RA, CC+RA, cc+EL, as defined in Table \ref{tab:tab8}. Low-mass galaxies: galaxies with $-19 < M_r < -16$~mag from the Fornax cluster catalogue (FCC, \citealt{ferguson:1989}; based on the magnitude conversions of \citealt{weinmann:2011}). Massive galaxies: galaxies with $M_r \leq -19$~mag from the FCC. Each massive galaxy is represented by a circle with three times its isophotal diameter at $\mu_B = 25$~mag~arcsec$^{-2}$, $3\,\rm{d_{25}}$ (we used the extinction-corrected values for d$_{25}$, obtained from HyperLEDA). The two brightest galaxies are NGC~1399 in the centre and NGC~1404 to the south-east from it.}
    \label{fig:fig9}
\end{figure*}

In Table~\ref{tab:tab8} we defined alternative subsamples, where we set the cut for the core concentration to a higher value in order to increase the number of objects in the subsamples with low core concentration. We show the spatial distributions of the alternative subsamples in Fig.~\ref{fig:fig9}, and summarize the KS test results in Table~\ref{tab:tab9}, in analogy to Section~\ref{sect:sect4.2}.

\begin{figure*}
	\includegraphics[width=0.79\textwidth]{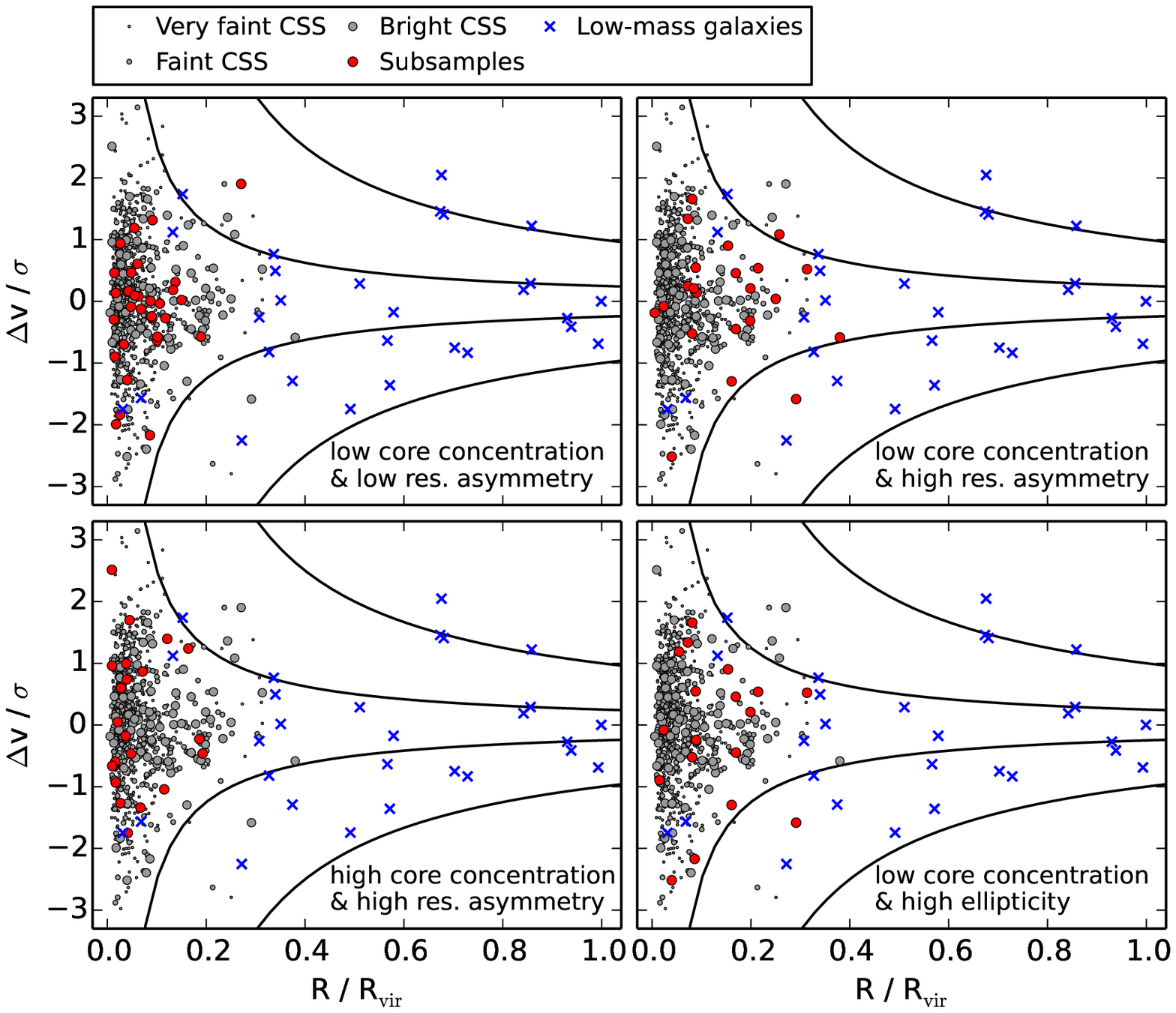}
    \caption{Phase-space distribution of compact stellar systems and low-mass galaxies in the Fornax cluster. Same as Fig.~\ref{fig:fig7}, but illustrating the alternative subsamples. Very faint CSS: compact objects with $m_{\mathrm{Ve}} \geq 21.5$~mag (not part of our working sample). Faint CSS: compact objects with $20.6 \leq m_{\mathrm{Ve}} < 21.5$~mag. Bright CSS: compact objects with $m_{\mathrm{Ve}} < 20.6$~mag. Subsamples: cc+ra, cc+RA, CC+RA, cc+EL, as defined in Table \ref{tab:tab5}. Low-mass galaxies: galaxies with $-19 < M_r < -16$~mag from the FCC (\citealt{ferguson:1989}; based on the magnitude conversions of \citealt{weinmann:2011}). $\Delta v$ is the relative velocity of an object with respect to the cluster mean velocity (1460~km~s$^{-1}$). We denote $\sigma$ as the cluster velocity dispersion (324~km~s$^{-1}$), $R$ as the cluster-centric distance, and $R_{\mathrm{vir}}$ as its virial radius ($2.5\degr$, see Section~\ref{sect:sect4.3}). The mean velocity and dispersion were calculated from all compact stellar systems and FCC galaxies within a cluster-centric distance of 1.0\degr. The solid lines correspond to caustic lines of constant $(\Delta v / \sigma)~\times~(R/R_{\mathrm{vir}})$ at $\pm$0.1 and $\pm$0.4, respectively.}
    \label{fig:fig10}
\end{figure*}

Fig.~\ref{fig:fig10} shows the distribution of the alternative subsamples in phase-space. We include the velocity dispersions of the alternative subsamples in Table~\ref{tab:tab10}.

\begin{table*}
 \caption{Parameter ranges for the alternative subsamples cc+ra, cc+RA, CC+RA and cc+EL. Compared to the definition in Table~\ref{tab:tab5}, the cut for the core concentration is set to a higher value. The residual asymmetry and ellipticity cuts remain unchanged. For each alternative subsample we give the fraction of objects in the respective magnitude range and the overlap fractions with the artificial objects.}
 \label{tab:tab8}
 \begin{tabular}{lllllll}
  \hline
  Alternative & Alternative cuts for & Objects & Art. objects & Alternative cuts for & Objects & Art. objects\\
  subsample & $m_{\mathrm{Ve}} < 20.0$~mag & (per cent) & (per cent) & $20.0 \leq m_{\mathrm{Ve}} < 20.6$~mag & (per cent) & (per cent)\\
  \hline
  cc+ra & cc < 0.80 and ra < 0.5 & 53.1 & 50.3 & cc < 0.87 and ra < 0.55 & 26.3 & 24.0\\
  cc+RA & cc < 0.80 and ra > 0.5 & 12.5 & \phantom{0}0.3 & cc < 0.87 and ra > 0.55 & 31.6 & \phantom{0}1.4\\
  CC+RA & cc > 0.80 and ra > 0.5 & 21.9 & \phantom{0}5.5 & cc > 0.87 and ra > 0.55 & 24.6 & \phantom{0}2.2\\
  cc+EL & cc < 0.80 and el\, > 0.2 & 15.6 & \phantom{0}1.5 & cc < 0.87 and el\, > 0.22 & 22.8 & \phantom{0}4.6\\
  \hline
 \end{tabular}
\end{table*}

\begin{table}
 \caption{Alternative subsamples: KS test probabilities (percentage) for the null hypothesis that two subsamples have the same cluster-centric distance distribution. In the last row the distributions of the individual subsamples are compared to the respective other compact objects with different parameters in the same magnitude range with $m_{\mathrm{Ve}} < 20.6$~mag.}
  \label{tab:tab9}
 \begin{tabular}{lllll}
  \hline
  Alternative & cc+ra & cc+RA & CC+RA & cc+EL\\
  subsample\\
  \hline
  cc+ra & 100.0 & \phantom{00}0.3 & \phantom{0}19.2 & \phantom{00}5.1\\
  cc+RA & & 100.0 & \phantom{00}0.0 & \phantom{0}97.4\\
  CC+RA & & & 100.0 & \phantom{00}0.4\\
  cc+EL & & & & 100.0\\
  respective & \phantom{0}12.2 & \phantom{00}0.1 & \phantom{00}0.8 &\phantom{00}4.0\\
  other CSS &&&&\\
  \hline
 \end{tabular}
\end{table}

\begin{table}
 \caption{Velocity dispersion for the alternative subsamples. The velocity dispersion of each subsample is calculated as standard deviation within a cluster-centric distance of $R \leq 1.0\degr$  ($\sigma_{\mathrm{tot}}$), $R \leq 0.4\degr$ ($\sigma_{\mathrm{in}}$) or $0.4 < R \leq 1.0\degr$ ($\sigma_{\mathrm{out}}$), based on the velocities given in Table~\ref{tab:tab1}. $N_{\mathrm{obj}}$ corresponds to the number of objects from the respective subsample in the inner ($N_{\mathrm{obj,in}}$) or outer ($N_{\mathrm{obj,out}}$) cluster region. The velocity dispersion is given in km~s$^{-1}$. $1.0\degr$ corresponds to $0.346$ and $0.4\degr$ to $0.138$~Mpc at the distance of Fornax.}
 \label{tab:tab10}
 \begin{tabular}{llllll}
  \hline
  Alternative & $\sigma_{\mathrm{tot}}$ & $\sigma_{\mathrm{in}}$ & $N_{\mathrm{obj,in}}$ & $\sigma_{\mathrm{out}}$ & $N_{\mathrm{obj,out}}$\\
  subsample\\
  \hline
  cc+ra & 284 & 265 & \phantom{0}30 & 401 & \phantom{0}2\\
  cc+RA & 303 & 340 & \phantom{0}11 & 252 & 11\\
  CC+RA & 359 & 374 & \phantom{0}18 & 244 & \phantom{0}3\\
  cc+EL & 375 & 427 & \phantom{0}11 & 269 & \phantom{0}7\\
  \hline
 \end{tabular}
\end{table}


\bsp	
\label{lastpage}
\end{document}